\documentclass[12pt]{iopart}
\usepackage{graphicx}
\usepackage{color}
\usepackage{amssymb}
\usepackage{float}
\def\
crpar{\mbox{$\not\!\partial$}}

\def\tab{\mbox{}\hspace{1.0cm}}

\def\XXint#1#2#3{{\setbox0=\hbox{$#1{#2#3}{\int}$}
     \vcenter{\hbox{$#2#3$}}\kern-.5\wd0}}

\newcommand{\la}{\label}
\newcommand{\be}{\begin{equation}}
\newcommand{\ee}{\end{equation}}
\newcommand{\bea}{\begin{eqnarray}}
\newcommand{\eea}{\end{eqnarray}}

\newcommand{\ba}{\begin{align}}
\newcommand{\ea}{\end{align}}

\def\XXint#1#2#3{{\setbox0=\hbox{$#1{#2#3}{\int}$}
     \vcenter{\hbox{$#2#3$}}\kern-.5\wd0}}

\begin{document}

\title[On Duality, emergence of Calogero family of models in external potentials, Solitons and Field theory
 \ldots]{Duality in a hyperbolic interaction model integrable even in a strong confinement: Multi-soliton solutions and field theory
 }

\author{Aritra Kumar Gon
}
\address{Indian Institute of Technology Madras, Chennai-600036, India}
\address{International Centre for Theoretical Sciences, Tata Institute of Fundamental Research, Bangalore - 560089, India}

\author{Manas Kulkarni
}
\address{International Centre for Theoretical Sciences, Tata Institute of Fundamental Research, Bangalore - 560089, India}

%
%

%

\begin{abstract}
Models that remain integrable even in confining potentials are extremely rare and almost non-existent. Here, we consider a one-dimensional hyperbolic interaction model, which we call as the Hyperbolic Calogero (HC) model. This  is classically integrable even in confining potentials (which have box-like shapes). We present a first-order formulation of the HC model in an external confining potential. Using the rich property of duality, we find multi-soliton solutions of this confined integrable model. Absence of solitons correspond to the equilibrium solution of the model. We demonstrate the dynamics of multi-soliton solutions via brute-force numerical simulations. We studied the physics of soliton collisions and quenches using numerical simulations. We have examined the motion of dual complex variables and found an analytic expression for the time period in a certain limit. We give the field theory description of this model and find the background solution (absence of solitons) analytically in the large-N limit. Analytical expressions of soliton solutions have been obtained in the absence of external confining potential. Our work is of importance to understand the general features of trapped interacting particles that remain classically integrable and can be of relevance to the collective behaviour of trapped cold atomic gases as well.


\end{abstract}
\maketitle
\tableofcontents
\markboth{}{}
\newpage
\section{Introduction}

The rational Calogero model (with and without an external Harmonic trap) and its various generalizations is one of the most well studied integrable system in physics and mathematics \cite{calogero1969ground,calogero1975exactly,calogero1969_1,calogero1971solution, moser1976three,polychronakos2006physics,polychronakos1992exchange}. In its traditional form, it describes $N$ identical non-relativistic particles in one dimension, interacting through two-body inverse-square potentials in the presence of an external harmonic potential \cite{calogero1971solution,mainmanas,CalogeroMosermodel}. This model and its various extensions appear in many branches of physics and mathematics and has connections and relevance to fractional statistics, fluid mechanics, spin chains, gauge theories, matrix models to name a few 
 \cite{olshanetsky1981classical,polychronakos2006physics,jevicki1980quantum,sakita1985quantum,
polychronakos1995waves,jevicki376nonperturbative,hikami1993integrable,
minahan1993integrable,polychronakos1993lattice,polychronakos1994exact}.  As a result, it has been studied extensively \cite{sutherland2004beautiful, olshanetsky1983quantum, olshanetsky1981classical,perelomov1990integrable}. The rational Calogero-Moser model is a model with power-law interaction where every particle is coupled to every other particle. This is therefore considered as a relatively long-ranged model although, in 1D, it might have few properties similar to that of relatively short-ranged models \cite{sutherland1975exact}. 
It was also shown that models with inverse square interactions in 1D, for some physical quantities, have similarity with short-ranged models \cite{kulkarni2011cold}. \\
%
%

Given that, in most physical realizations, one has confined particles interacting with each other over some length scales, it would be of importance to study systems that remain integrable  \cite{polychronakos1992quartic,perelomov1990integrable} even in confined potentials. For e.g., a recent breakthrough in cold atomic systems has been the realization of an almost uniform gas of atoms confined in a box-like potential \cite{gaunt2013bose}. One of the main challenges is to find integrable models that continue to remain integrable even when they are confined in external potentials. 
Hence, we are looking for integrable systems with two properties: (i) Inter-particle interactions where every particle interacts with every other particle over some length-scale and (ii) Strong confinement after some legnth scale. Therefore, we study the model described below which has inter-particle interactions over some length scale, strong external confinement and the rich property of classical integrability. In addition to these properties, the model we consider has the rich property of duality and a field theory formulation. \\

In this paper, we investigate the behaviour of a classical system in a confining potential, interacting through an inverse square sine-hyperbolic potential. This can be viewed as a generalization of the rational Calogero-Moser model which is now periodic on the imaginary line. This is called the Hyperbolic Calogero (HC) Model.\\

The  general form of Hamiltonian for the HC model reads,
\bea
{\cal H}=\sum_{i=1}^{N}\left[\frac{p_{i}^{2}}{2}+V(x_i)+\sum_{i,\,j\neq i}^{N}\frac{1}{2L^{2}}\left(\frac{g^{2}}{\sinh^{2}\left(\frac{x_{i}-x_{j}}{L}\right)}\right)\right]
\label{gham}
\eea
where
\bea
\label{vx}
V(x_{i})=a_{1}\cosh\left(\frac{2x_{i}}{L}\right)+b_{1}\sinh\left(\frac{2x_{i}}{L}\right)+a_{2}&&\cosh\left(\frac{4x_{i}}{L}\right)+b_{2}\sinh\left(\frac{4x_{i}}{L}\right)\nonumber \\
\eea
where $x_{j}$ are the coordinates of the particles, $p_{j}$ are their canonical momenta, $L$ is a length scale associated with the model, $g$ is the coupling constant and $N$ is the number of particles. We take the mass of the particles to be unity.
The above model is classically integrable \cite{polychronakos1992new}.
Apart from the interaction between particles, another basic difference between rational and hyperbolic model is in the structure of the external potential. In the hyperbolic case, the external potential is essentially flat within a certain region from the origin and rises steeply after that thus acting like a confinement for the particles moving in it. The size of the system turns out to be $L_c \sim L\sinh^{-1} \Big(\sqrt \frac{g N}{A}\Big)$ where $A$ is the strength of the external confining potential (to be discussed later). Particles are confined within this length. We find that the length of the system essentially scales logarithmically (hence, very slowly) with number of particles which is very different from the rational model where particles spread out as their number increases (length in rational case scales as $\sqrt{N}$). This creates a major difference in the particle density profile.\\

In Section \ref{dhc}, we formulate the HC Hamiltonian with an external potential from a first order equation involving dual variables. 
We formulate the initial position and velocity distributions of particles for obtaining soliton solutions. This means finding a special set of initial conditions $\{ x_i (0), p_i(0)\}$ where $i=1,2... N$ such that when the particles move under the influence of the Hamiltonian, the density profile of the particles is a robust moving soliton. We find multi-solitons in this integrable system. This is done by solving the damping equation that we explain later. Then, we studied the dynamics of the particles for these special set of initial conditions by solving differential equations numerically. We examined the integrals of motion which provides information about the integrability of the system. We also checked the effects of two, three and four soliton collisions. We examined the effects of the various parameters on the background density (density without formation of any soliton). We have also checked the effects of quenching the parameters on the soliton motion. \\

In Section 3, we formulate the equations of motion for the dual variables and analyze their trajectories in the complex plane. We also check the integrals of motion. We have also examined the motion of the dual variable after quenching. We find an analytic form for the time period of oscillation of the soliton by computing periodicity of motion of the dual variables in the complex plane. \\

In Section 4, we  derive and study the  field theory formulation of this model under the continuum limit and present the corresponding soliton solutions in terms of meromorphic fields. For this discussion, the effects of the external potential are neglected as the effective length of the box was taken to be infinity. Here the particles are essentially replaced by a density field.
The integrability and other rich properties of the underlying particle systems suggest that the corresponding fluid mechanical equations are also integrable and point to the existence of soliton solutions for the HC field theory (without external potential). We find out an analytic form to represent the equilibrium density distribution for the background density profile as well as for soliton solutions. We find that the equilibrium density, i.e., the background density  (absence of solitons)  is similar to a hyperbolic version of the trigonometric equations that appear in the context of large-N gauge theories \cite{gross1993possible, wadia1980n}. 
 We also provide an analytic expression for soliton velocity and expressed its connection with the motion of the dual variable in the complex plane\\

Finally, in Section 5, we state our conclusions and provide directions of our future investigation.

\section{Dual Hyperbolic Calogero system and formation of the Hamiltonian}
\label{dhc}
In this section, we aim to formulate the first-order dual equations of motion for the dynamical system of particles of the confined HC model. 

To start with, we consider a system of $N$ particles with coordinates $x_i$,
$i = 1, \dots , N$, and $M$ dual-particles with coordinates $z_n$, $n = 1, \dots , M$, moving in the complex plane obeying the {\it first-order} equations of motion \cite{mainmanas},
\bea
\label{xdot}
\dot{x_{i}}-i\frac{A}{L}\sinh\left(\frac{2x_{i}}{L}\right)=-i\frac{g}{L}\sum_{j\neq i}^{N}&&\coth\left(\frac{x_{i}-x_{j}}{L}\right)\nonumber\\
&&+i\frac{g}{L}\sum_{n=1}^{M}\coth\left(\frac{x_{i}-z_{n}}{L}\right)    
\eea
\bea 
\label{zdot}
\dot{z_{n}}-i\frac{A}{L}\sinh\left(\frac{2z_{n}}{L}\right)=i\frac{g}{L}\sum_{m\neq n}^{M}&&\coth\left(\frac{z_{n}-z_{m}}{L}\right)\nonumber\\
&&-i\frac{g}{L}\sum_{i=1}^{N}\coth\left(\frac{z_{n}-x_{i}}{L}\right)
\eea
These are a set of two first order coupled differential equations describing the motion of $N$ values of $x_i$ and $M$ values of $z_n$. The dynamics are fully described by the initial values of these $M$+$N$ variables. One can show that if the initial values of $x_i$ are chosen to be real,  they remain so for all future times. Using this formalism we can map the motion of $N$ particles moving in real axis to motion of $M$ dual variables moving in the complex plane. The number of dual variables are not related to the numbers of particles in the real axis, i.e. this formalism is valid even for $M<N$. 
Remarkably the second order equations completely decouple from each other. They are of the following form,
\bea
\label{xddot}
\ddot{x_{i}}&=&-\frac{2A^{2}}{L^{3}}\sinh\left(\frac{2x_{i}}{L}\right)\cosh\left(\frac{2x_{i}}{L}\right)+\frac{2Ag}{L^{3}}(N-M-1)\sinh\left(\frac{2x_{i}}{L}\right)\nonumber\\
&&+\frac{2g^{2}}{L^{3}}\sum_{j\neq i}^{N}\left(\frac{\cosh\left(\frac{x_{i}-x_{j}}{L}\right)}{\sinh^{3}\left(\frac{x_{i}-x_{j}}{L}\right)}\right)\textbf{\hspace{1.5 in}$j=1.....N$ }
\eea
\bea
\label{zddot}
\ddot{z_{n}}&=&-\frac{2A^{2}}{L^{3}}\sinh\left(\frac{2z_{n}}{L}\right)\cosh\left(\frac{2z_{n}}{L}\right)+\frac{2Ag}{L^{3}}(N-M+1)\sinh\left(\frac{2z_{n}}{L}\right)\nonumber\\
&&+\frac{2g^{2}}{L^{3}}\sum_{m\neq n}^{M}\left(\frac{\cosh\left(\frac{z_{n}-z_{m}}{L}\right)}{\sinh^{3}\left(\frac{z_{n}-z_{m}}{L}\right)}\right)\textbf{\hspace{1.5 in}$n=1.....M$ } 
\eea
Once the initial condition of positions and conjugate momenta for the particles (on the real line) are obtained using first order Eq.~\ref{xdot} we can study the dynamics of those particles from Eq.~\ref{xddot} without requiring further information about the dual variables.\\
The Hamiltonian corresponding to Eq.~\ref{xddot} is
\bea
 {\cal H} &=& \sum_{i=1}^{N}\left(\frac{p_i^{2}}{2}+\frac{A^{2}}{2L^{2}}\sinh^{2}\left(\frac{2x_{i}}{L}\right)-\frac{Ag}{L^{2}}(N-M-1)\cosh\left(\frac{2x_{i}}{L}\right)\right)\nonumber\\
   &+& \sum_{i,\,j\neq i}^{N}\frac{1}{2L^{2}}\left(\frac{g^{2}}{\sinh^{2}\left(\frac{x_{i}-x_{j}}{L}\right)}\right)
     \la{HyCM}
  \la{Hsquare}
\eea

\subsection{Multi-Soliton Solutions}
Solitons are excitations (solitary pulse like structures) that are formed due to the collective motion of the particles. These solitons do not disperse or break down as the particles move with time. The resulting density profile (collective behaviour) is a robust excitation which does not break or disperse. 
Soliton solutions belong to a very special set of initial conditions in the space of initial conditions where the motion of all the particles shows a coherent behaviour. This occurs due to the delicate interplay between non-linearity, non-locality and dispersive effects. This is very sensitive to system parameters. We will further analyse its structure and behaviour in later sections.\\

We will now present the way by which the initial conditions for soliton solutions can be obtained in general. Note that finding soliton solutions means restricting the space of initial conditions of $N$ values of $x_i$ and $p_i$. 
By equating the imaginary part of Eq.~\ref{xdot}, we get the following equation,
\bea
\label{fixsol}
-\frac{A}{L}\sinh\left(\frac{2x_{i}}{L}\right)&=&-\frac{g}{L}\sum_{j\neq i,\atop j=1}^{N}\coth\left(\frac{x_{i}-x_{j}}{L}\right)\nonumber\\
&&+\frac{g}{L}Re\sum_{n=1}^{M}\left[\coth\left(\frac{x_{i}-z_{n}}{L}\right)\right]\textbf{\hspace{0.5 in}$i=1......N$} 
\eea
Equating the real part of Eq.~\ref{xdot}, we get the conjugate momenta ($p_i \equiv \dot{x}_i$),
\bea
p_{i}=\frac{g}{L}Im\left[\sum_{n=1}^{M}\coth\left(\frac{x_{i}-z_{n}}{L}\right)\right]
\textbf{\hspace{0.5 in}$i=1......N$} 
\label{fixvsol}
\eea
The fixed points of these set of $N$ equations in Eq.~\ref{fixsol} give the equilibrium position of particles for obtaining soliton solution. Essentially, at that point  $ \dot{x_{i}}=0$ for all  $i=1,2... N$. We can equivalently write Eq.~\ref{fixsol} as,
\bea
\label{dudx}
\frac{\partial U}{\partial x_{i}}=0\textbf{\hspace{0.5 in}$i=1......N$} 
\eea
where,
\bea
\frac{\partial U}{\partial x_{i}}= \frac{A}{L}\sinh\left(\frac{2x_{i}}{L}\right)+\frac{g}{L}\sum_{j\neq i,\atop j=1}^{N}&&\coth\left(\frac{x_{i}-x_{j}}{L}\right)\nonumber\\
&&-\frac{g}{L}Re\sum_{m\neq n}^{M}\left[\coth\left(\frac{x_{i}-z_{n}}{L}\right)\right]
\eea 
Therefore, we can form a damping equation \cite{polymanas17} for a chosen set of $z$ values,
\bea
\label{damp}
\dot{x}_{i}=-\gamma\frac{\partial U}{\partial x_{i}}
\eea
where $\gamma$ can be considered as some damping coefficient. This is actually the numerical way we employ for  solving Eq.~\ref{fixsol} for obtaining solutions corresponding to a local minimum. The damping equation in principle slides the particles towards the minimum of the above potential. The damping acts like a viscous force which slides the particles towards equilibrium position. This will finally give us the special set of $\{ x_i(t=0)\}$ and the special set $\{ p_i (t=0) \}$ is obtained from Eq.~\ref{fixvsol}.

\subsection{Background}

\begin{figure}[H]
\centering
\includegraphics[width=0.9\linewidth]{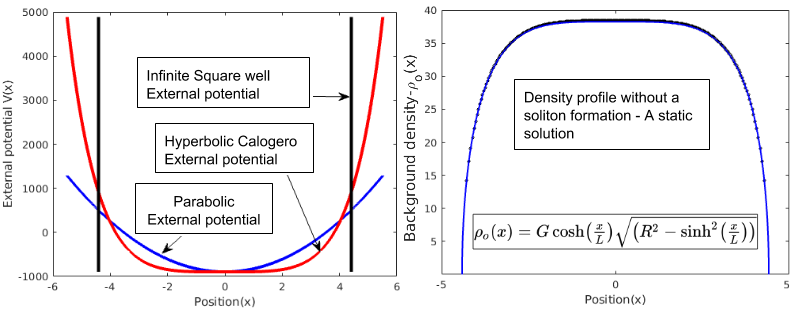}
\caption{(Left) Various confining potentials (Right) Density profile without the soliton $\rho_0(x)$. There is no dual variable, M=0, L=5, Number of particles N = 300, g=0.5. A=150. Points represent brute-force numerics and line represents our analytical expression. }
\label{fig:bkg301}
\end{figure}

The background constitutes the position of particles which corresponds to $M=0$ (no solitonic excitations). It gives us a static solution. From Eq.~\ref{fixvsol}, it is clear that when there is no dual variable, we have $N$ values of $p_j$ to be equal to $0$. Hence, this is called the static solution. No soliton formation occurs. The particles just sit in their equilibrium position. For this situation, Eq.~\ref{fixsol} becomes,
\bea
\label{background}
\frac{A}{L}\sinh\left(\frac{2x_{i}}{L}\right)&=&\frac{g}{L}\sum_{j\neq i\atop j=1}^{N}\coth\left(\frac{x_{i}-x_{j}}{L}\right)\textbf{\hspace{0.5 in}$i=1......N$}
\eea
Solving the corresponding damping equation, we get the positions of $j$ particles. We then plot the density of the particles as a function of position (see Fig.~\ref{fig:bkg301}). Note that, just for plotting purpose in Fig.~\ref{fig:bkg301}, the density $\rho(x)$ is defined as the inter-particle distance and the position index is taken as the mean position between the corresponding two particles. This should not be confused with the classical density field $\rho(x)$ that we introduce later in the field theory section (Sec.~\ref{cft}). 

\subsection{Relationship of background solutions with generalized Log gas}
\label{glog}

It is interesting to note that the equilibrium solutions of the classical HC model (Eq.~\ref{Hsquare} with $M=0$) which is given by Eq.~\ref{background} is also the minimum energy configuration of a generalized version of Log gas given by, 
\begin{equation}
V_{\log}  =\frac{A}{2}\sum_{i=1}^N\cosh \bigg( \frac{2 x_i}{L}\bigg) - \frac{g}{L}\sum_{i \neq j}^N\frac{1}{2} \log \Big| \sin \bigg(\frac{ x_i - x_j }{L} \bigg) \Big|
\label{vglog}
\end{equation}
Although, there has been a great deal of work on connections between traditional Log gas and Random Matrix Theory \cite{forrester2010log,o2010gaussian,gustavsson2005gaussian,majumdar2014top} and their relation to Calogero-Moser systems \cite{CalogeroMosermodel}, to the best of our knowledge, little is understood about the relationship between classical HC model, the generalized version of Log gas and Random Matrix Theory.  It is to be noted that trigonometric version of the above generalized Log gas (Eq.~\ref{vglog}) effectively appear in the context of large-N gauge theories \cite{gross1993possible,wadia1980n}. 


We get a plateau like graph where the density $\rho_0$ is essentially constant throughout the characteristic length  of the box and falls sharply  after that. This replicates the flatness of the external potential within the box-like potential. We were able to obtain the exact functional form of this curve which will be discussed in the field theory section (Sec~4.). We will also show the variation of density with system parameters in Sec~4.


\subsection{One soliton solutions}

\begin{figure}[H]
\centering
\includegraphics[width=0.7\linewidth]{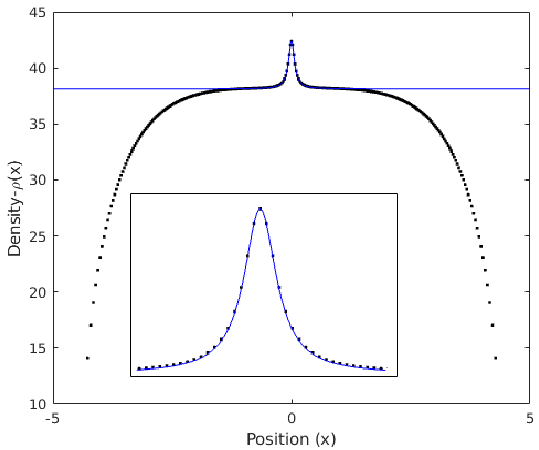}
\caption{One soliton density profile for 300 particles. Since, its a one soliton solution, only one dual variable was needed. The blue line represent our ansatz for the soliton (Eq.~\ref{anzsol}). We used L=5, g=0.5, A=150. The dual variable was at $z(t=0) = 0.078356i$ in the complex plane}
\label{fig:density101_L5}
\end{figure}

We now consider the case where $M=1$. This corresponds to one dual variable moving in the complex plane. For this case Eq.~\ref{fixsol} becomes,
\bea
\frac{A}{L}\sinh\left(\frac{2x_{i}}{L}\right)=-\frac{g}{L}\sum_{i,\,j\neq i}^{N}\coth\left(x_{i}-x_{j}\right)+\frac{g}{L}Re\left[\coth\left(\frac{x_{i}-z}{L}\right)\right]  
\label{solone}
\eea
The momenta are given by,
\bea
p_{i}=\frac{g}{L}Im\left[\coth\left(\frac{x_{i}-z}{L}\right)\right]\textbf{\hspace{0.5 in}$i=1......N$}
\eea
After obtaining the initial conditions, we get the corresponding density profile. The density profile is plotted here by calculating inverse of inter-particle distance (see Fig.~\ref{fig:density101_L5}). 

We can observe a bump at the origin. This is essentially the soliton. This forms due to the dual variable which in principle acts like an attractor of particles, thus increasing the density near the origin. We have observed that the height of the bump depends on the distance of the dual variable from the real axis. The lesser the distance of the dual particle from real axis, more is height of the resulting soliton. We have made an ansatz for the analytic form of the soliton which we will discuss later in the field theory section (Sec. 4).\\

\begin{figure}[H]
\centering
\includegraphics[width=0.7\linewidth]{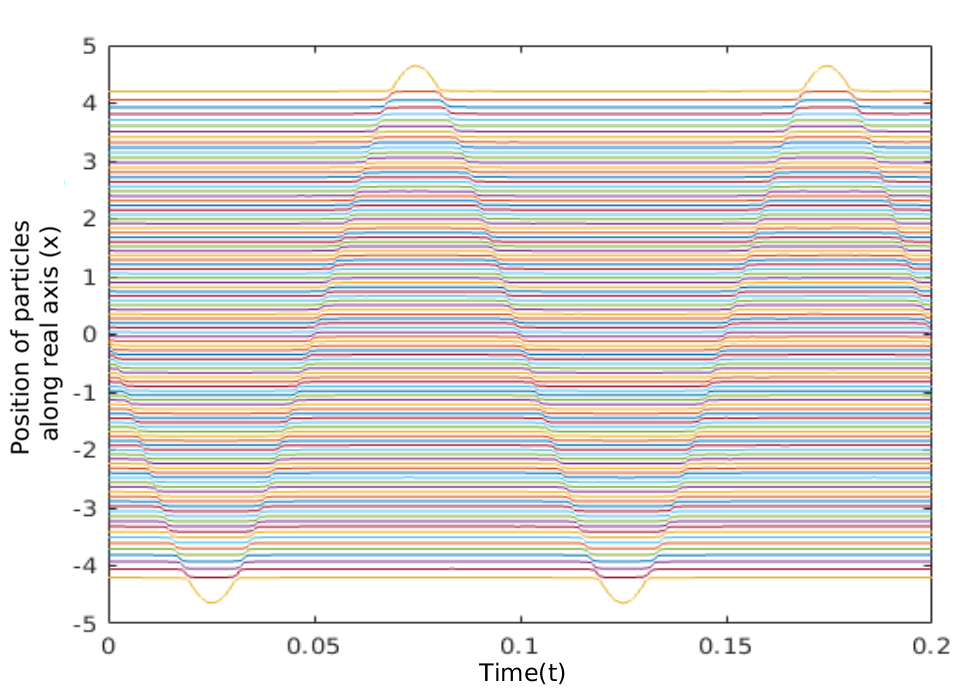}
\caption{World lines for $101$ particles (we haven't shown the 300 particles plot for better clarity). Here g=0.5, A=50.5. The dual variable was at $z(t=0) =0.078356i$ in the complex plane. Each line belongs to an individual particle. Therefore, the particle trajectories of 101 particles are plotted here. The wave-like curve is the result of coherent motion and corresponds to a single soliton. As we can see, the particles are always bounded within the box.}
\label{fig:101}
\end{figure}

\begin{figure}[H]
	\centering
	\includegraphics[width=0.8\linewidth]{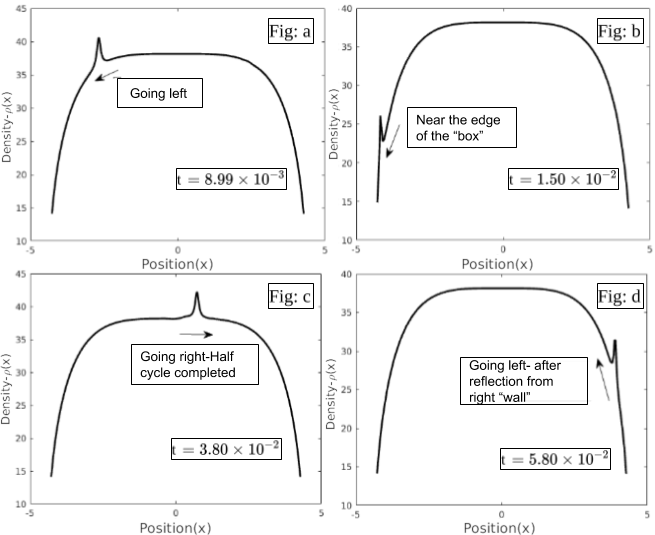}
	\caption{Time evolution of the density profile. There is a single soliton which oscillates from one end of the box-like potential to the other. The soliton does not disintegrate as time evolves, i.e, it remains fully robust. Here A=150 , N=300, g=0.5. The dual variable at $z(t=0)= 0.078356i$}
	\label{fig:one}
\end{figure}

 Using numerical simulations, we have observed the evolution of particles using the second order differential equations, Eq.~\ref{xddot}. The particle trajectories are plotted in Fig.~\ref{fig:101} (world-lines). We also examined the evolution of soliton density with time (see Fig.~\ref{fig:one}). 

We observe a coherent motion of the particles as time evolves. As a result there is a perfect wave-like motion which can be seen very clearly, though individually they move only a little from their equilibrium positions. This is the analog of the Newton's Cradle \cite{kinoshita2006quantum}. The soliton maintains its form and does not break or disperse which is exactly what we expect from a soliton evolution. This kind of robust evolution is highly non-trivial and involves a delicate balance between various effects such as non-linearity, non-locality and dispersion.  
We also checked the integrability of the system, which is essentially examining the integrals of motion. We checked the $1^{st}$ integral which is the energy of the system and also the $2^{nd}$ integral of motion. These quantities were conserved with very high accuracy for very long times.\\

Soliton stability analysis is a subject of great interest and we plan to address this for the HC model in future. Our numerics indicate that by slightly perturbing the soliton solution, we still  retain robust behaviour at least for several many time periods, i.e., for a very considerable long time \footnote{We thank E. Bogomolny for pointing us to this interesting problem for the Hyperbolic Calogero model. }.

\subsection{Multi-soliton evolutions (two ,three and four soliton solutions)}
It is important to note that we can find multi-soliton solutions in the confined HC model by exploiting the $M<N$ duality. The existence of multi-soliton solutions is a consequence of classical integrability of the confined HC model. In this section, we find the multi-soliton solutions. Further, here we check the effects of multi-soliton collisions. For $M=2$, $M=3$ and $M=4$ there are interactions between the dual variables too. As a result, we expect to see interesting dynamics in the complex plane as well. The motion of dual variables will be discussed in later sections (Sec.~3).
\begin{figure}[H]
	\centering
	\includegraphics[width=0.8\linewidth]{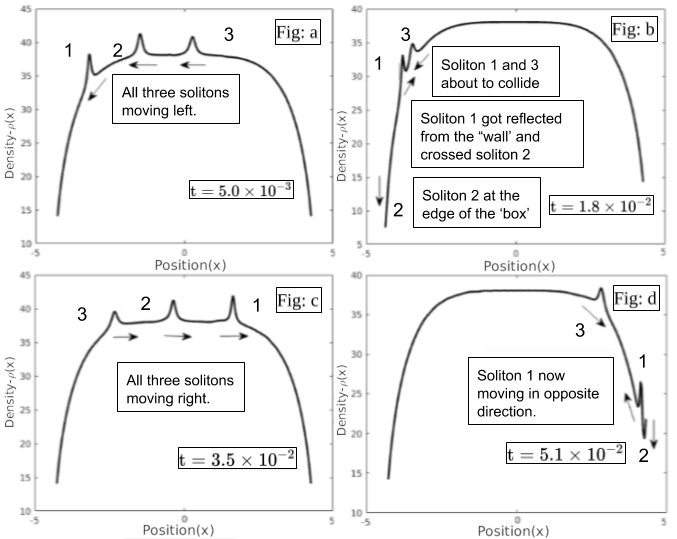}
	\caption{Three soliton density profiles at different times. The solitons are of different heights and pass through each other. Note that the densities are constructed as inverse of inter-particle distance. Here M=3, L=5, N=300, g=0.5 and A=150. The three dual variables are at $z_1(t=0)= 1.75+0.118356i$, $z_2(t=0)= 0.098356i$ and $z_3(t=0)= -1.75+0.078356i$}
	\label{fig:3diff}
\end{figure}

\begin{figure}[H]
	\centering
	\includegraphics[width=0.92\linewidth]{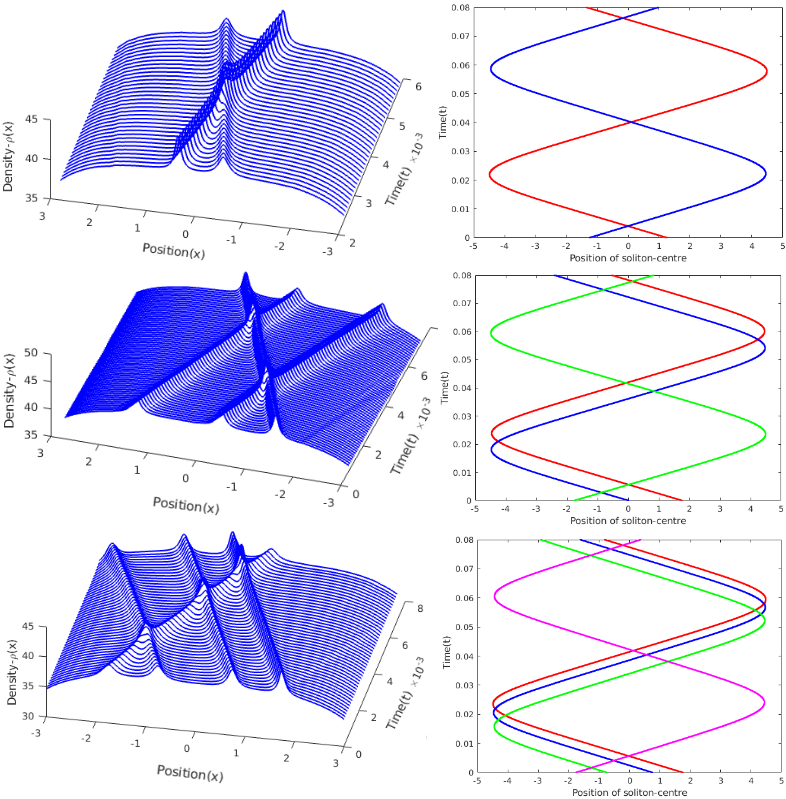}
	\caption{Soliton train diagram describing the evolution of two solitons (top left), three solitons (middle left) and four solitons (bottom left). Initially for two solitons, the dual variables were at $z_1(t=0) = 1.25+0.078356i$ and $z_2(t=0) = -1. 25-0.138356i$, for three solitons the dual variables were at $z_1(t=0) =1.75+0.118356i$, $z_2(t=0) =0.098356i$ and $z_3(t=0) =-1.75-0.078356i$ and for the four solitons the dual variables were at $z_1(t=0) =1.75+0.078356i$, $z_2(t=0) =0.75+0.108356i$, $z_3(t=0) = -0.75+0.138356i$ and $z_4(t=0) = -1.75-0.158356i$. The solitons passes through each other without getting destroyed. Adjacent to that, we have plotted the time evolution of the soliton guiding centres for two solitons (top right), three solitons (middle right) and four solitons (bottom right).}
	\label{fig:density_dynamics}
\end{figure}

In Fig.~\ref{fig:3diff}, we show three soliton solutions and their evolutions. As can be seen, they pass through each other which is a consequence of integrable nature of the solitons.  In Fig.~\ref{fig:density_dynamics}, we show the soliton train diagram for two, three and four solitons (left panel in Fig.~\ref{fig:density_dynamics}). It is important to note that the position of real part of the dual variables determine the position of the solitons, the  magnitude of the imaginary parts determine the height of the solitons (greater the magnitude, lesser the height) and finally the signs of the imaginary part dictate in which direction the solitons will move (if the sign is negative, the soliton moves right and if the sign is positive, the soliton moves to the left). The right panel in  Fig.~\ref{fig:density_dynamics} shows the motion of the guiding centre of the solitons. It can be seen that the centres of the solitons pass through each other and also bounce off the walls. 
 
\subsection{Quenching}

In this section we see the effects of suddenly changing a system parameter such as the coupling constant $g$. We mainly examine the changes in soliton evolution due to this quench \cite{franchini2015universal,franchini2016hydrodynamics}.\\
\begin{figure}[H]
\centering
\includegraphics[width=0.8\linewidth]{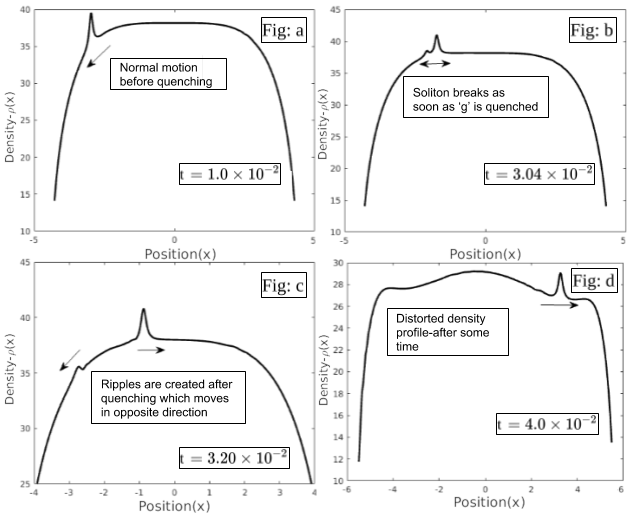}
\caption{Here N=300, A=150 and the dual variable was $z(t=0)= 0.078356i$. (a) We see that before quenching the soliton oscillates normally. (b) When the parameter g is quenched from 0.5 to 0.8 we see that the soliton just breaks/splits (c) Ripples move in the opposite direction. (d) After some long time, the density profile is distorted with some ripples.}
\label{fig:testc1}
\end{figure}

\begin{figure}[H]
\centering
\includegraphics[width=0.8\linewidth]{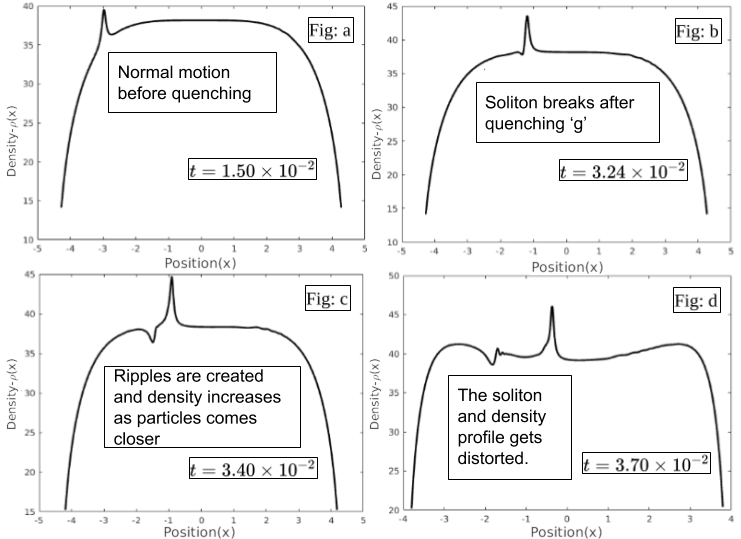} 
\caption{Here N=300, A=150 and the dual variable was $z(t=0)= 0.078356i$. (a) We see that before quenching the soliton oscillates normally. (b) When the parameter g is quenched from 0.5 to 0.25, we see that the soliton just breaks at that moment. (c) As we progress in time, we see that dips are formed. (d)  After some long time, the density gets distorted. }
\label{fig:testc}
\end{figure}

In Fig.~\ref{fig:testc1}, we observe that the soliton breaks down and ripples are formed which bounces back and forth inside the box-like potential (see caption in Fig.~\ref{fig:testc1}). In  Fig.~\ref{fig:testc}, we see that, when the coupling constant is decreased the particles repel each other with a lesser strength and so they can come much closer to each other. On the other hand, if the coupling constant is increased (Fig.~\ref{fig:testc1}), the exact opposite phenomena occurs. There is a discontinuity in the energy during quenching which is expected as the interaction as well as the external potential depends on $g$, but the new energy (post-quench) remains constant.

\section{Dynamics of Dual particles in complex plane}
\label{ddc}
In this section, we focus on the motion of dual variables corresponding to one and two-soliton solution. 
We find the connection between the motion of dual variables and the real Calogero particles. We also check for the energy conservation for the dual system. We find an analytic solution of a single dual variable motion in the small $y$ limit and the time period of motion. We also examine the effect of quenching on the dual variable. 

\subsection{Single Dual variable corresponding to one soliton solution}
\begin{figure}[H]
\centering
\includegraphics[width=0.7\linewidth]{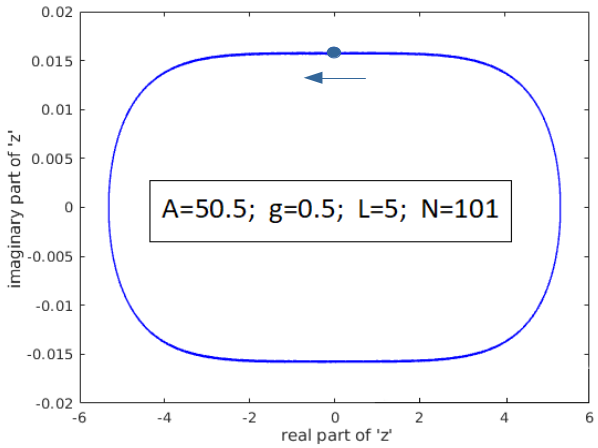}
\caption{Single dual variable in complex plane. The initial position is shown in the figure with the blue dot. One complete revolution of this dual variable corresponds to one complete oscillation of the soliton density profile. This figure describes the movement of dual particle corresponding to dynamics shown in the world line Fig. 3 }
\label{fig:cplane_1sol_L5_t=3_g1_A31_n31b}
\end{figure}

Once the initial condition for the position of real variables $\{x_i (t=0)\}$ are obtained from the damping equation, i.e., Eq.~\ref{damp}, we can find the initial momenta of the dual variables using Eq.~\ref{zdot}. For a single $z$, Eq.~\ref{zdot} takes the form,
\bea
\label{1z dot}
\dot{z}-i\frac{A}{L}\sinh\left(\frac{2z}{L}\right)=-i\frac{g}{L}\sum_{j=i}^{N}\coth\left(\frac{z-x_{j}}{L}\right) 
\eea
We choose the initial position of the dual variable on the imaginary axis, i.e. $Re[z]=0$. In the corresponding density plot, we see that the initial position of the soliton is centred at the origin. Once the initial position and momenta are determined, the evolution is governed by Eq.~\ref{zddot}.  
\bea
\label{1zddot}
\label{single_z}
\ddot{z}=-\frac{2A^{2}}{L^{3}}\sinh\left(\frac{2z}{L}\right)\cosh\left(\frac{2z}{L}\right)+\frac{2Ag}{L^{3}}N\sinh\left(\frac{2z}{L}\right) 
\eea
We solve the differential equation numerically to get Fig.~\ref{fig:cplane_1sol_L5_t=3_g1_A31_n31b} which forms a rectangle-like trajectory in the complex plane.

We see that the motion of the dual variable is also confined within a box-like potential. It forms a closed trajectory in the complex plane which is like a smeared rectangle. Thus, it is periodic and we expect it to have a definite time period. As the dual variable moves, it drags the soliton with it. This gives us insight that one complete cycle of the dual variable corresponds to one complete oscillation of the soliton. Thus we expect to find exact same time period for them. We will discuss it in a later section.
\subsection{Two Dual variables corresponding to two soliton solution}
We repeat the same mechanism for two dual variables instead of one. In this case, there will be an  interaction term in the governing equation. Therefore, the analogous equations to Eq.~\ref{1z dot} and Eq.~\ref{1zddot} are, 
\bea
\dot{z_{1}}=i\frac{A}{L}\sinh\left(\frac{2z_{1}}{L}\right)+i\frac{g}{L}\coth\left(\frac{z_{1}-z_{2}}{L}\right)-i\frac{g}{L}\sum_{i=1}^{N}\coth\left(\frac{z_{1}-x_{i}}{L}\right)\nonumber\\
\dot{z_{2}}=i\frac{A}{L}\sinh\left(\frac{2z_{2}}{L}\right)+i\frac{g}{L}\coth\left(\frac{z_{2}-z_{1}}{L}\right)-i\frac{g}{L}\sum_{i=1}^{N}\coth\left(\frac{z_{2}-x_{i}}{L}\right)
\eea
and 
\\
\\
\bea
\ddot{z_{1}}=-\frac{2A^{2}}{L^{3}}\sinh\left(\frac{2z_{1}}{L}\right)\cosh\left(\frac{2z_{1}}{L}\right)+\frac{2Ag}{L^{3}}&&(N-1)\sinh\left(\frac{2z_{1}}{L}\right)\nonumber\\
&&+\frac{2g^{2}}{L^{3}}\left(\frac{\cosh\left(\frac{z_{1}-z_{2}}{L}\right)}{\sinh^{3}\left(\frac{z_{1}-z_{2}}{L}\right)}\right)\nonumber
\eea
\bea
\ddot{z_{2}}=-\frac{2A^{2}}{L^{3}}\sinh\left(\frac{2z_{1}}{L}\right)\cosh\left(\frac{2z_{2}}{L}\right)+\frac{2Ag}{L^{3}}&&(N-1)\sinh\left(\frac{2z_{2}}{L}\right)\nonumber\\
&&+\frac{2g^{2}}{L^{3}}\left(\frac{\cosh\left(\frac{z_{2}-z_{1}}{L}\right)}{\sinh^{3}\left(\frac{z_{2}-z_{1}}{L}\right)}\right)
\eea
\begin{figure}[H]
\centering
\includegraphics[width=0.75\linewidth]{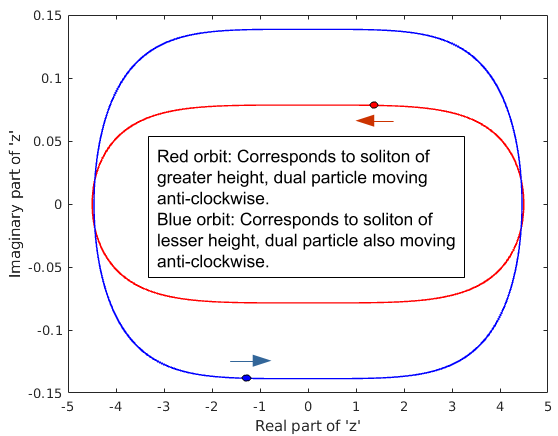}
\caption{Two dual variables at different $Im[z]$. These correspond to two solitons of different heights.
This shows why the solitons pass through each other. We see that the two dual variables seem to interact very weakly and move fairly independently in their own orbits. So the corresponding solitons also move independently without interacting, i.e., the solitons just pass through. Here A=150, N=300, g=0.5. The above plot is over several time periods. The initial positions of the dual variables are shown in the figure with the blue and red dot.}
\label{fig:diff_heightz}
\end{figure}
Fig.~\ref{fig:diff_heightz} shows the trajectory of the two dual variables. We see that they essentially do not change their trajectory. This is also reflected in the motion of the soliton. We have discussed earlier that the solitons pass through each other unhindered. It is very non-trivial to understand the reason just by observing the real particles. The motion of the dual variables on the other hand gives a more transparent intuition into the interaction process. We have stated earlier that the dual variables drag the solitons with them. 

\subsection{Effects of quenching on Dual variable}
We have seen  that due to quenching the soliton breaks and the particles either spread out a bit or get contracted inward. This result is reflected in the motion of the dual variable.
\begin{figure}
\centering
\includegraphics[width=0.9\linewidth]{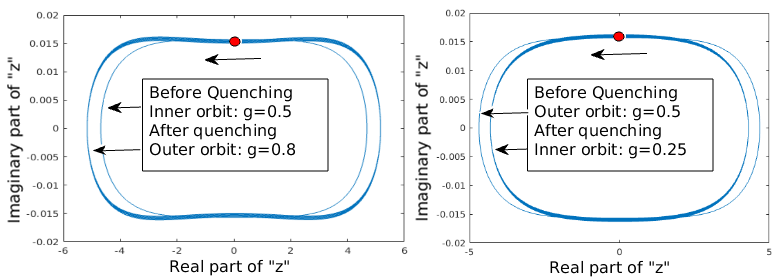}
\caption{The two figures shows the effect on dual variable due to quenching. Though the dual variable maintains the same kind of trajectory but the mapping no longer is maintained, so the evolution is not at all like a soliton evolution. Here A=150, N=300. The initial position is shown in the figures with the red dot.\\
	(Left) When g is increased the particles repel each other more and hence they spread out. Hence the dual variable move in a larger orbit\\
	(Right) When g is decreased the repulsion decreases hence they shrink to smaller region. Correspondingly the dual variable move in a smaller orbit}
\label{fig:quench_z_101_g05_08d}
\end{figure}
We observe that after quenching, the dual variable either shifts to a larger closed curve or a smaller one (depending on the nature of the quench). If the value of coupling constant $g$ is increased the particle spreads out, so the dual variables also move in a bigger orbit and if $g$ is decreased the particles get contracted and so the $z$ moves in a smaller orbit.\\

We note that although the real Calogero particles upon quench break into waves, the dual variables still seem to have an ordered motion.  So, it is evident that the mapping in general breaks as it is no longer a soliton evolution. 
Since the post-quench dynamics is no longer a soliton evolution, a single-z dual variable is not sufficient to describe the collective behaviour of Calogero particles.

\subsection{Analytic solution of Dual variable in small y limit}

In this section, we solve the equation of motion for $z$ to find an analytic form of the solution. Our aim is to then find a formula for the periodicity of its orbit. Then, we can actually match this with our simulation results. We need to make the small-y approximation which essentially means that the imaginary part of $z$ is very small in comparison to the length of the box. This assumption is justified. Indeed, for having a proper soliton formation, the dual variable must be quite close to the x-axis. This is true for our model and system parameters. The equation of motion for a single dual variable is given by Eq.~\ref{single_z}.\\
Writing the real part and imaginary part of $\cosh(x+iy)$ and then taking small-y limit, i.e,  $\cos(y)\rightarrow1$, we get (for the real part of $z$), the following, 
\bea
\ddot{x}=-\frac{A^{2}}{L^{3}}\sinh\left(\frac{4x}{L}\right)+\frac{2Ag}{L^{3}}N\sinh\left(\frac{2x}{L}\right)
\eea
To get the periodicity, we can look at time evolution of the above equation. The solution of the above equation is, 
\bea
x=L\left[\tan^{-1}\left(c_{1}\mathit{\mathrm{JacobiSN}(c_{2}it,m)}\right)\right]
\eea  
Above, $c_1$ and $c_2$, are determined by the initial conditions, $x(0)$ and $\dot{x}(0)$.  
This \textit{JacobiSN} function is one of the Jacobi elliptic functions. This function is periodic in nature. The periodicity here depends only on the parameter $m$ which in turns depends of the initial position, initial velocity and system parameters such as $g$ and $A$. The exact dependence of $m$ on these parameters are not explicit but we know the periodicity as a function of $m$ which is,
\bea
\label{timeP}
T=\frac{\mathrm{Re}\left[4\mathrm{EllipticK(1-m)}\right]}{c_{2}}
\eea 
where $\mathrm{EllipticK(x)}$ is the complete elliptic integral of $1^{st}$ kind.
\bea
\mathrm{EllipticK(m)}=\int_{0}^{\frac{\pi}{2}}\left(\frac{1}{\sqrt{1-m\sin^{2}(\theta)}}\right)d\theta
\eea
This result matches extremely well with our simulation results.  
\section{Collective field theory formulation}
\label{cft}
In this section, we derive the collective field theory for the HC model. 
Under the continuum limit, we choose $N\rightarrow \infty$. Further we neglect the effects of confining potential during the formulation of the Hamiltonian.  Here, the position and momentum of the individual variables get replaced by continuous density field $\rho(x)$ and velocity field $v(x)$. We aim to formulate the Hamiltonian as a function of these fields. Then, we form the continuity equation and Euler equation and finally we try to establish the ansatz for the analytic form for several things such as background solutions,  soliton solutions etc;
\subsection{Formulation of Hamiltonian}
The general Hamiltonian without the external potential is of the form \cite{polychronakos1992new},
\bea
{\cal H}=\sum_{i=1}^{N}\left\{ p_{i}^{2}+\sum_{j\neq i}^{N}\frac{g^{2}}{2L^{2}}\left(\frac{1}{\sinh^{2}\left(\frac{x_{i}-x_{j}}{L}\right)}\right)\right\} 
\eea
In the continuous limit, we replace the position of individual variables by a position function such that $x(j)=x_j$. This is assumed to be a smooth function. The derivative of this function is related to the density field as,
\bea
\label{xprimej}
x'(j)=\frac{dx}{dj}=\frac{1}{\rho(x)}
\eea
We can show that, under the continuum limit, we get (see Appendix B for details), 
\bea
\lim_{N\rightarrow\infty}\sum_{j\neq i}^{N}\frac{g^{2}}{2L^{2}}\left(\frac{1}{\sinh^{2}\left(\frac{x(i)-x(j)}{L}\right)}\right)=g\left(\pi\rho^{\mathrm{H}}-\partial_{x}log\sqrt{\rho(x)}\right)^{2}
\eea
Then the Hamiltonian becomes,
\bea
{\cal H}=\int dx\rho(x)\left[\frac{v^{2}}{2}+\frac{1}{2}\left(\pi g\rho^{\mathrm{H}}-g\partial_{x}log\sqrt{\rho(x)}\right)^{2}\right]+const
\eea
where $\rho(x)^\mathrm{H}$ is the Hilbert transform of $\rho(x)$ and is defined as in \cite{stone,polymanas17},
\bea
\label{hilbert}
\rho(x)^{\mathrm{H}}=\frac{1}{\pi L}P\left\{ \int_{-\infty}^{\infty}\left[\rho(\tau)\coth\left(\frac{\tau-x}{L}\right)d\tau\right]\right\}
\eea
The detailed discussions about the Hilbert transform is presented in Appendix C.
\subsection{Analytic form of the background density}
\subsubsection{Analytic form from background equations}
In this section, we will provide an analytic form of the density profile without the formation of the soliton. First, we form an equivalent field theory version of the background equation, Eq.~\ref{background}. So, we have,
\bea
\label{startbackground}
\frac{A}{L}\sinh\left(\frac{2x_{i}}{L}\right)=\frac{g}{L}\sum_{j\neq i}^{N}\coth\left(\frac{x_{i}-x_{j}}{L}\right)
\eea
Also, we know,
\bea
\rho(x)=\sum_{i=1}^{N}\delta(x-x_{i})
\eea
So multiplying the above equation on both sides of Eq.~\ref{startbackground} we have,
\bea
\frac{A}{L}\sum_{i=1}^{N}\delta(x-x_{i})\sinh\left(\frac{2x_{i}}{L}\right)=\frac{g}{L}\sum_{i=1}^{N}\sum_{j\neq i}^{N}\delta(x-x_{i})\coth\left(\frac{x_{i}-x_{j}}{L}\right)
\eea

\bea
\frac{A}{L}\rho(x)\sinh\left(\frac{2x}{L}\right)=\frac{g}{L}\lim_{N\rightarrow\infty}\sum_{j\neq i}^{N}\rho(x)\coth\left(\frac{x-x_{j}}{L}\right)
\eea
Now from Appendix B (Part 1) we know,
\bea
\frac{g}{L}\lim_{N\rightarrow\infty}\sum_{j\neq i}^{N}\rho(x)\coth\left(\frac{x-x_{j}}{L}\right)=g\rho(x)\left(\frac{\partial}{\partial x}\ln\sqrt{\rho(x)}-\pi\rho(x)^{\mathrm{H}}\right)
\eea
So we have,
\bea
\frac{A}{L}\rho(x)\sinh\left(\frac{2x}{L}\right)=g\rho(x)\left(\frac{\partial}{\partial x}\ln\sqrt{\rho(x)}-\pi\rho(x)^{\mathrm{H}}\right)
\eea
which gives, 
\bea
\label{Fieldbackground}
\frac{A}{L}\sinh\left(\frac{2x}{L}\right)+g\pi\rho(x)^{\mathrm{H}}=g\frac{\partial}{\partial x}\ln\sqrt{\rho(x)}
\label{brho}
\eea
We shall argue later that in the large-N limit, we can ignore the $\log$ term (i.e., the right hand side of Eq.~\ref{brho}). Keeping this in mind, we propose an ansatz for the functional form of the background density,
\bea
\rho_{o}(x)&&=G\cosh\left(\frac{x}{L}\right)\sqrt{R^2-\sinh^{2}\left(\frac{x}{L}\right)}\textbf{\hspace{0.5in}}(|x|<L\sinh^{-1}R)\nonumber\\
&&=0\textbf{\hspace{2.6in}}otherwise
\eea
where the parameters $G$ and $R$ will be fixed later. We will now prove that this ansatz satisfies Eq.~\ref{Fieldbackground}. We have, 
\bea
\rho_{0}(x)^{\mathrm{H}}=P\left(\int_{-\infty}^{\infty}\frac{G}{\pi L}\cosh\left(\frac{\tau}{L}\right)\sqrt{R^{2}-\sinh^{2}\left(\frac{\tau}{L}\right)}\coth\left(\frac{\tau-x}{L}\right)d\tau\right)
\eea
We now split the integrals into two parts. One, in which the integrand contains only the regular part of the integral and other which contains the singular part and so that it needs to be evaluated in the principal value sense. Therefore we have,
\bea
\label{bkndI1I2}
&&\rho_{0}(x)^{\mathrm{H}}=\int_{-L_{1}}^{L_{1}}\frac{G}{\pi L}\cosh\left(\frac{\tau}{L}\right)\left[\sinh^{2}\left(\frac{x}{L}\right)-\sinh^{2}\left(\frac{\tau}{L}\right)\right]\frac{\coth\left(\frac{\tau-x}{L}\right)}{\sqrt{R^{2}-\sinh^{2}\left(\frac{\tau}{L}\right)}}d\tau\nonumber\\
&&+P\left(\int_{-L_{1}}^{L_{1}}\frac{G}{\pi L}\cosh\left(\frac{\tau}{L}\right)\left(R^{2}-\sinh^{2}\left(\frac{x}{L}\right)\right)\frac{\coth\left(\frac{\tau-x}{L}\right)}{\sqrt{R^{2}-\sinh^{2}\left(\frac{\tau}{L}\right)}}d\tau\right)
\eea
where $L_1=L\sinh^{-1}R$. So,
\bea
I_1=\int_{-L_{1}}^{L_{1}}\frac{G}{2\pi L}\cosh\left(\frac{\tau}{L}\right)\Bigg[\cosh\left(\frac{2x}{L}\right)&&-\cosh\left(\frac{2\tau}{L}\right)\Bigg]
\frac{\coth\left(\frac{\tau-x}{L}\right)}{\sqrt{R^{2}-\sinh^{2}\left(\frac{\tau}{L}\right)}}d\tau\nonumber\\
\eea
\bea
=\int_{-L_{1}}^{L_{1}}-\frac{G}{2\pi L}\cosh\left(\frac{\tau}{L}\right)\frac{\sinh\left(\frac{2x}{L}\right)+\sinh\left(\frac{2\tau}{L}\right)}{\sqrt{R^{2}-\sinh^{2}\left(\frac{\tau}{L}\right)}}d\tau
\eea
\bea
=\int_{-L_{1}}^{L_{1}}-\frac{G}{2\pi L}\cosh\left(\frac{\tau}{L}\right)&&\frac{\sinh\left(\frac{2x}{L}\right)}{\sqrt{R^{2}-\sinh^{2}\left(\frac{\tau}{L}\right)}}d\tau\nonumber\\
&&-\int_{-L_{1}}^{L_{1}}\frac{G}{2\pi L}\cosh\left(\frac{\tau}{L}\right)\frac{\sinh\left(\frac{2\tau}{L}\right)}{\sqrt{R^{2}-\sinh^{2}\left(\frac{\tau}{L}\right)}}d\tau
\eea
Therefore we again have two separate integrations. The integral,
\bea
\int_{-L_{1}}^{L_{1}}\frac{G}{2\pi L}\cosh\left(\frac{\tau}{L}\right)\frac{\sinh\left(\frac{2\tau}{L}\right)}{\sqrt{R^{2}-\sinh^{2}\left(\frac{\tau}{L}\right)}}d\tau=0
\eea
as the integrand is an odd function and the limits of integration is symmetric about $\tau$. So we need to evaluate the other integral (denoted by $I_3$). 
\bea
I_3=\int_{-L_{1}}^{L_{1}}-\frac{G}{2\pi L}\cosh\left(\frac{\tau}{L}\right)\frac{\sinh\left(\frac{2x}{L}\right)}{\sqrt{R^{2}-\sinh^{2}\left(\frac{\tau}{L}\right)}}d\tau
\eea
Changing the variable $\sinh(\frac{\tau}{L})=z$ we have $\cosh(\frac{\tau}{L})d\tau=Ldz$ we have,
\bea
I_3&=&\int_{-R}^{R}-\frac{G}{2\pi L}\frac{\sinh\left(\frac{2x}{L}\right)}{\sqrt{R^{2}-z^{2}}}Ldz \nonumber\\
&=&\frac{G}{2}\sinh\left(\frac{2x}{L}\right)
\label{i3eq}
\eea
Now, we need to evaluate the other part (i.e., the principal part) of Eq.~\ref{bkndI1I2} which is,

\bea
I_2=\frac{G}{\pi L}\left[R^{2}-\sinh^{2}\left(\frac{x}{L}\right)\right]P\left(\int_{-L_{1}}^{L_{1}}\cosh\left(\frac{\tau}{L}\right)\frac{\coth\left(\frac{\tau-x}{L}\right)}{\sqrt{R^{2}-\sinh^{2}\left(\frac{\tau}{L}\right)}}d\tau\right)
\eea
We have performed this integration using brute force numerical technique  and found it to be equal to 0 (with machine precision). Hence, we can safely say that this principal value integral is equal to 0.\\

So in order to satisfy Eq.~\ref{brho} (but without the log term) we must have $G=\left(\frac{2A}{\pi Lg}\right)$ (since it has to obey Eq.~\ref{i3eq}). So, our proposed ansatz for the background density function stands,
\bea
\rho_{o}(x)=\left(\frac{2A}{\pi Lg}\right)\cosh\left(\frac{x}{L}\right)\sqrt{R^{2}-\sinh^{2}\left(\frac{x}{L}\right)}
\eea
One way to justify ignoring the Log term is as follows. If we take Eq.~\ref{brho} and rescale as follows, $\rho(x)= N \tilde \rho(x)$, then we get, 

\begin{equation}
\frac{A}{L}\sinh\left(\frac{2x}{L}\right)+g\pi N \tilde{\rho}(x)^{\mathrm{H}}=g\frac{\partial}{\partial x}\ln\sqrt{\tilde{\rho}(x)}
\end{equation}
Considering that $A\sim O(N)$, we notice that the Log term is $1/N$ suppressed which therefore can be neglected in large-N limit. The irrelevance of the Log term can also be seen in an alternate way in the next discussion (Sec. \ref{aft}) on field theory description of the Hamiltonian. It is interesting to note that a trigonometric version of Eq.~\ref{startbackground} (discrete) and of Eq.~\ref{Fieldbackground} (continuum and without $\log$ term) appears in the context of Gross-Witten-Wadia phase transition in large-N gauge theories \cite{gross1993possible, wadia1980n} which was analysed using the approach of Brezin, Itzykson, Parisi, and Zuber \cite{brezin1978brezin}.

In order to establish a strong evidence to the above analytical result, we have matched our ansatz with numerical simulation of the density profiles with various parameters (see Fig.~\ref{fig:tr}). We find perfect agreement between the analyical expression and brute-force numerics. 
By imposing, normalization condition, i.e., $\int_{-L\sinh^{-1}(R)}^{L\sinh^{-1}(R)} \rho(x) dx = N$, we get  $R=\sqrt{\frac{gN}{A}}$.
So the final expression for the background density stands to be, 
\bea
\label{finalexprho}
\rho_{o}(x)=\left(\frac{2A}{\pi Lg}\right)\cosh\left(\frac{x}{L}\right)\sqrt{\left(\frac{gN}{A}\right)-\sinh^{2}\left(\frac{x}{L}\right)}
\eea 
From the above equation we observe that,
\bea
\rho_{o}(0)=\left(\frac{2}{\pi L}\sqrt{\frac{AN}{g}}\right)
\eea

\subsubsection{Analytical form from field theory of the Hamiltonian}
\label{aft}
Without the need of an ansatz, the analytical solutions for the background density can be obtained, via writing down the large-N field theory of the Hamiltonian Eq.~\ref{Hsquare} and then using variational principle.  We can derive the field theory at large $N$ to be \cite{polymanas17}, 
\bea
 {\cal H} =\int_{-\infty}^{+\infty}\Bigg[\frac{1}{2}\rho v^2+\frac{g^2 \pi^2 \rho^3}{6}&+&\frac{A^{2}}{2L^{2}}\sinh^{2}\left(\frac{2x}{L}\right)\rho(x)\nonumber \\ &-&\frac{AgN}{L^{2}}\cosh\left(\frac{2x}{L}\right)\rho(x)\Bigg]\nonumber\\
  \la{HsquareF}
\eea
Taking the variational derivative w.r.t $\rho$, i.e., $\frac{\delta H}{\delta \rho}$ along with a chemical potential $\mu$, gives, 
\begin{equation}
\frac{g^2 \pi^2 \rho^2}{2}+\frac{A^{2}}{2L^{2}}\sinh^{2}\left(\frac{2x}{L}\right)-\frac{AgN}{L^{2}}\cosh\left(\frac{2x}{L}\right) = \mu
\end{equation}
which immediately gives, 
\bea
\label{rhoxf}
\rho(x) = \frac{\sqrt{2}}{\pi g}\sqrt{\mu -\frac{A^{2}}{2L^{2}}\sinh^{2}\left(\frac{2x}{L}\right)+\frac{AgN}{L^{2}}\cosh\left(\frac{2x}{L}\right)}
\eea

By putting  $\mu = \frac{g N A}{L^2}$ which sets the limits of integration to $\pm \beta$ where $\beta = L \sinh ^{-1} \Big[ \sqrt{\frac{g N}{A}}\Big]$, we satisfy the normalization,  $\int_{-\beta}^{+\beta} \rho(x)dx = N $. Plugging, this expression for $\mu$  back into Eq.~\ref{rhoxf}, after some algebra gives exactly the expression Eq. \ref{finalexprho}. Therefore, we arrive at the background density without the need for an ansatz. 

\subsubsection{Some observation on the background density:}
\label{transition}

From Eq.~\ref{finalexprho}, we can analyse how the shape of background density changes with different parameter values. The first derivative of that equation gives,
\bea
\rho_{o}'(x)=\left(\frac{2A}{\pi Lg}\right) \frac{\left(gN-A\cosh\left(\frac{2x}{L}\right)\right)\sinh\left(\frac{x}{L}\right)}{AL\sqrt{\frac{gN}{A}-\sinh^{2}\left(\frac{x}{L}\right)}}
\eea
Therefore $\rho_{o}'(x)=0$ for $x=0$ and $\cosh(\frac{2x}{L})=\frac{gN}{A}$. But, the minimum value of $\cosh(\frac{2x}{L})$ is 1 (since it is a $\cosh$ function). Therefore, for $\frac{gN}{A}<1$ we have only one extrema  (i.e., $x=0$). For $\frac{gN}{A}>1$ we get two more extrema values (in addition to $x=0$). They can be found to be, 
\bea
x=\pm
L\cosh^{-1}\sqrt{\frac{1}{2}\left(1+\frac{gN}{A}\right)}
\label{xtwo}
\eea

\begin{figure}[H]
\centering
\includegraphics[width=0.32\linewidth]{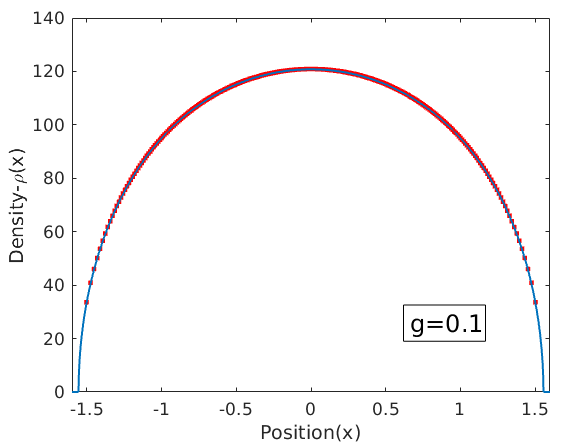}
\includegraphics[width=0.32\linewidth]{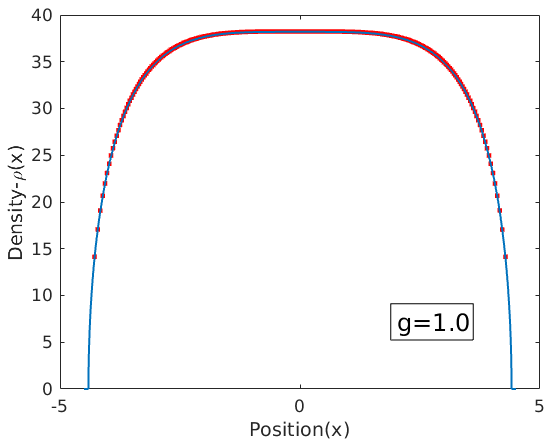}
\includegraphics[width=0.32\linewidth]{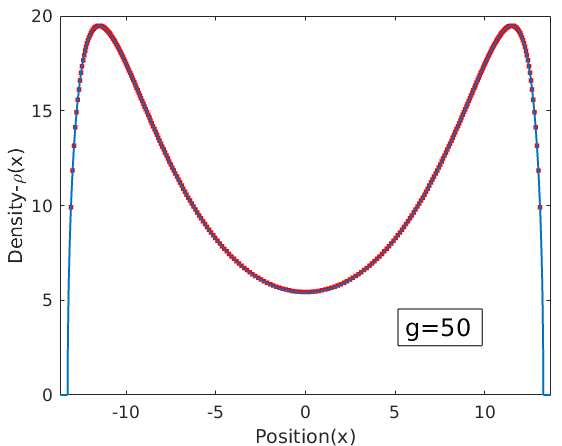}
\caption{Background density in the three regimes with $N=300$, $A=300$, $L=5$. The solid line represents the analytical results using Eq. \ref{finalexprho} and the red dots represent the brute force numerical data using Eq.~\ref{damp}. (Left) The case of $g<1$ (g=0.1) where we have one maxima and a dome-like density profile. (Middle) The case when g=1 which has an inflection point (table-top structure). (Right) This case when $g>1$ (g=50), which has one minima and two maxima. In this case, the particles gets pushed out to the edge of the box. Thus we can say that there is a transition in the density profile at $g=1$.}
\label{fig:tr}
\end{figure}

Taking the second derivative we find that we get only one maxima at $x=0$ for $\frac{gN}{A}<1$. For the other case of $\frac{gN}{A}>1$, we get a minima at $x=0$ and two maxima at $x=\pm
L\cosh^{-1}\sqrt{\frac{1}{2}\left(1+\frac{gN}{A}\right)}$.
In the case of $\frac{gN}{A}=1$ which, for Eq.~\ref{xtwo}, implies $L\cosh^{-1}\sqrt{\frac{1}{2}\left(1+\frac{gN}{A}\right)}=0$, the  three extrema points coincide at $x=0$ and we have an inflection point. So, fixing $A$ and $N$ to some value we can observe a transition in the functional form of the background density as we change the coupling constant $g$. At $\frac{gN}{A}=1$ there is a change in the curvature of the  background density function which is also evident from Fig.~\ref{fig:tr}.  This analysis is closely connected to the Gross-Witten-Wadia phase transition in large-N gauge theories \cite{gross1993possible, wadia1980n}. 

\begin{figure}[H]
\centering
\includegraphics[width=0.32\linewidth]{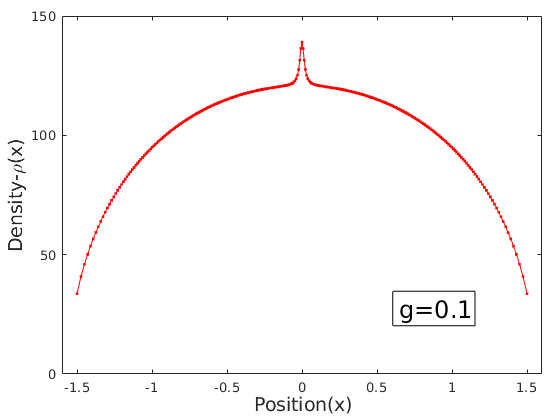}
\includegraphics[width=0.32\linewidth]{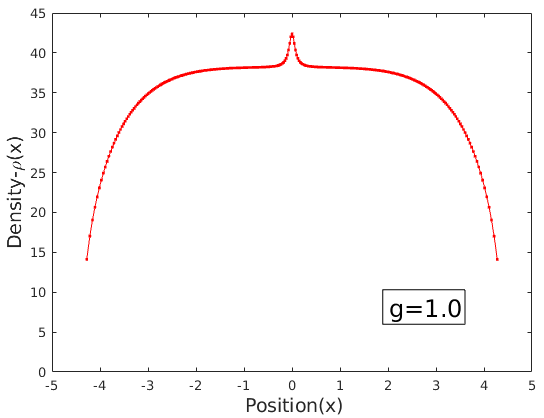}
\includegraphics[width=0.32\linewidth]{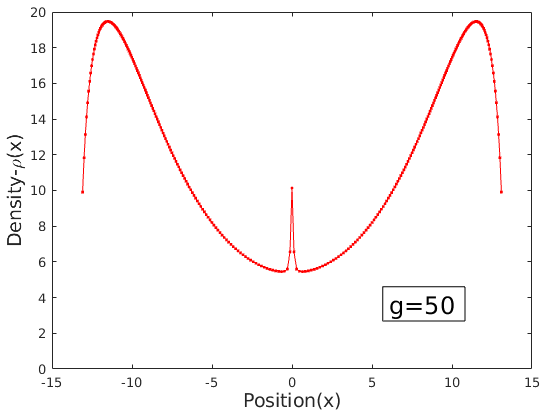}
\caption{Soliton profiles in the three regimes with $N=300$, $A=300$, $L=5$ via the brute force numerical data using Eq.~\ref{solone}. (Left) The case of $g<1$ (g=0.1) and $z_1=0.018356i$. (Middle) The case when g=1 and $z_2=0.078356i$. (Right) This case when $g>1$ (g=50) and $z_3=0.018356i$}
\label{fig:trsol}
\end{figure}

Fig.~\ref{fig:trsol} shows the one soliton solution sitting on top of the background in all three cases, $g>1,g=1$ and $g<1$. The corresponding time evolution will be a solitonic excitation moving on top of such non-trivial backgrounds.


\subsection{Analytic form of the density field and soliton solutions in Field Theory}
As an equivalent form of the those dual equations we introduce two meromorphic fields $U^{+}(x)$ and $U^{-}(x)$ \cite{mainmanas}, 
\bea
\label{uplus}
U^{+}(x)=\frac{ig}{L}\sum_{n=1}^{M}\coth\left(\frac{x-z_{n}}{L}\right)
\eea
\bea
\label{uminus}
U^{-}(x)=-\frac{ig}{L}\sum_{j=1}^{N}\coth\left(\frac{x-x_{j}}{L}\right)
\eea  
These equations are still not defined in the continuum limit. $U^{-}(x)$ has poles on the real axis only while $U^{+}(x)$ has poles which are not on the real axis. We have 
\bea
\label{uminusxj}
U^{+}(x_{j})=p_{j}+\frac{ig}{L}\sum_{k\neq j,1}^{N}\coth\left(\frac{x_{j}-x_{k}}{L}\right)
\eea   
We introduce the corresponding particle density field 
\bea
\rho(x)=\sum_{j=1}^{N}\delta(x-x_{j})
\eea
Now we can represent Eq.~\ref{uminus} in the following parametric form,
\bea
\label{cauchy}
U^{-}(z)=-\frac{ig}{L}\int_{-\infty}^{\infty}  \rho(x) \coth\Big(\frac{z-x}{L}\Big)dx
\eea
The above equation is independent of the exact number of particles. So this equation can be extended even to the continuum limit, where of course the form of the density function will be different. Eq.~\ref{cauchy} is discontinuous on the real axis,
\bea
\lim_{\epsilon\rightarrow0}U^{-}(x_{0}-i\epsilon)&=&-\frac{ig}{L}\lim_{\epsilon\rightarrow0}\Bigg[\int_{-\infty}^{x{}_{0}-\epsilon}\rho(x)\coth\left(\frac{(x_{0}-i\epsilon)-x}{L}\right)dx\nonumber\\
&&+\int_{x_{0}+\epsilon}^{\infty}\rho(x)\coth\left(\frac{(x_{0}-i\epsilon)-x}{L}\right)dx\nonumber\\
&&+\int_{x_{0}-\epsilon}^{x_{0}+\epsilon}\rho(x)\coth\left(\frac{(x_{0}-i\epsilon)-x}{L}\right)dx\Bigg]
\eea
From the definition of principal value integral we have,
\bea
\label{u(x-e)}
 \lim_{\epsilon\rightarrow0}U^{-}(x_{0}-i\epsilon)&=&-\frac{ig}{L}\lim_{\epsilon\rightarrow0}\Bigg[P\left\{ \int_{-\infty}^{\infty}\rho(x)\coth\left(\frac{z-x}{L}\right)dx\right\} \Bigg|_{z=(x_{0}-i\epsilon)}\nonumber\\
&&+\int_{x_{0}-\epsilon}^{x_{0}+\epsilon}\rho(x)\coth\left(\frac{(x_{0}-i\epsilon)-x}{L}\right)dx\Bigg]
\eea
We will now simplify the second expression on the r.h.s of the above equation.
\\
Now we observe, 
\bea
\lim_{\epsilon\rightarrow0}\int_{x_{0}-\epsilon}^{x_{0}+\epsilon}\rho(x)&\coth&\left(\frac{x-(x_{0}-i\epsilon)}{L}\right)dx\nonumber\\
&&+\lim_{r\rightarrow0}\int_{c_{1}}\rho(z)\coth\left(\frac{z-(x_{0}-i\epsilon)}{L}\right)dz\nonumber\\
&&=\oint\rho(z)\coth\left(\frac{z-(x_{0}-i\epsilon)}{L}\right)dz
\eea
Here we choose a closed contour of semicircular shape of radius $r$ which closes in the upper half plane, consisting of curve $\mathcal{C}_1$ in Fig.~\ref{fig:contour_plot1} and the real axis joining the curve. We have shifted the singular point below the real axis and the contour closes in anticlockwise direction. Using residue theorem we have (see Appendix C).
\bea
\label{resi-final}
\oint\rho(z)\coth\left(\frac{z-(x_{0}-i\epsilon)}{L}\right)dz=0
\eea
Therefore,
\bea
 \lim_{\epsilon\rightarrow0}\int_{x_{0}-\epsilon}^{x_{0}+\epsilon}\rho(x)\coth\left(\frac{(x_{0}-i\epsilon)-x}{L}\right)dx=i\pi L\rho(x_{0})
\eea
So from Eq.~\ref{u(x-e)} we get the final expression as,
\bea
 \lim_{\epsilon\rightarrow0}U^{-}(x_{0}-i\epsilon)=-\frac{ig}{L}\left[-\pi L\rho(x_{0})^{\mathrm{H}}+i\pi L\rho(x_{0})\right]
\eea
Thus, we have, 
\bea
 U^{-}(x-i0)=\pi g\rho(x)+\mathit{i \pi g\rho(x)^{\mathrm{H}}}
\eea
Similarly we can shift the singular point upward by $\epsilon$ and then do our calculation in which case the contour integral, Eq.~\ref{resi-final} will produce $2\pi iL\rho(x_{0})$. So,
\bea
\lim_{\epsilon\rightarrow0} \int_{x_{0}-\epsilon}^{x_{0}+\epsilon}\rho(x)\coth\left(\frac{x-(x_{0}+i\epsilon)}{L}\right)dx+i\pi L\rho(x_{0})=2\pi iL\rho(x_{0})
\eea
which implies
\bea
\lim_{\epsilon\rightarrow0}\int_{x_{0}-\epsilon}^{x_{0}+\epsilon}\rho(x)\coth\left(\frac{(x_{0}+i\epsilon)-x}{L}\right)dx=-i\pi L\rho(x_{0})
\eea
So, finally we get,
\bea
U^{-}(x\mp i0)=\pm\pi g\rho(x)+\mathit{i\pi g\rho(x)^{\mathrm{H}}}
\eea 
We need to find the equivalent form of $U^{+}(x)$ in the continuum limit. From Eq.~\ref{uminusxj} we have,
\bea
\sum_{j=1}^{N}\delta(x-x_{j})U^{+}(x_{j})&=&\sum_{j=1}^{N}\delta(x-x_{j})p_{j}\nonumber\\
&&+\frac{ig}{L}\sum_{j=1}^{N}\sum_{k\neq j}^{N}\delta(x-x_{j})\coth\left(\frac{x_{j}-x_{k}}{L}\right)
\eea

\bea
\label{rhouplus}
\rho(x)U^{+}(x)=\rho(x)v(x)+i\frac{g}{L}\rho(x)\sum_{k\neq j}^{N}\coth\left(\frac{x-x_{k}}{L}\right)
\eea
where, we define $v(x)$ such that, 
\bea
\rho(x)v(x)=\sum_{j=1}^{N}\dot{x}_{j}\delta(x_{i}-x_{j})
\eea
and
\bea
\sum_{j=1}^{N}\delta(x-x_{j})U^{+}(x_{j})=\rho(x)U^{+}(x)
\eea
Next, we will go to the continuum limit, i.e., $N\rightarrow\infty$. As in the case of Hamiltonian, the positions of individual variables are replaced by an equivalent position functions $x=x(j)$, where j represents the $j^{th}$ particle. It is related to the density field by Eq.~\ref{xprimej}. Therefore Eq.~\ref{rhouplus} takes the form,
\bea
\rho(x)U^{+}(x)=\rho(x)v(x)+\frac{ig}{L}\lim_{N\rightarrow\infty}\sum_{k\neq j\atop k=-N}^{N}\rho(x)\coth\left(\frac{x(j)-x(k)}{L}\right)
\eea
We can show that (see Appendix B), 
\bea
\label{HlimitU(x)}
\lim_{N\rightarrow\infty}\sum_{k\neq j\atop k=-N}^{N}\coth\left(\frac{x(j)-x(k)}{L}\right)=L\frac{\partial}{\partial x}\ln\sqrt{\rho(x)}-\pi L\rho(x)^{\mathrm{H}}
\eea
where $\rho(x)^{\mathrm{H}}$ is defined in Eq.~\ref{hilbert}. Therefore, we have, 
\bea
\rho(x)U^{+}(x)=\rho(x)v(x)+i\frac{g}{L}\rho(x)\left(L\frac{\partial}{\partial x}\ln\sqrt{\rho(x)}-\pi L\rho(x)^{\mathrm{H}}\right)
\eea
Thus dividing by $\rho(x)$ we get
\bea
\label{fieldUplus}
U^{+}(x)=v(x)+ig\left(\frac{\partial}{\partial x}\ln\sqrt{\rho(x)}-\pi\rho(x)^{\mathrm{H}}\right)
\eea
As stated earlier the number of dual variables is independent of the number of real particles. So, under the continuum limit too we can find one soliton solutions.
From Eq.~\ref{uplus} and Eq.~\ref{fieldUplus} for $M=1$ we get,
\bea
\label{(U+fieldeqn)}
\frac{ig}{L}\coth\left(\frac{x-z}{L}\right)=-ig\left(\pi\rho^{H}(x)-\partial_{x}\log\sqrt{\rho(x)}\right)+v(x)
\eea
Therefore, equating the imaginary part of the equation, we get,
\bea
\label{ansatzcheck}
g\left(\pi\rho^{H}-\partial_{x}log\sqrt{\rho(x)}\right)=-\frac{g}{2L}\left[\coth\left(\frac{x-z}{L}\right)+\coth\left(\frac{x+z}{L}\right)\right]
\eea 
This equation is difficult to solve explicitly. We can guess a solution and then see whether it satisfies the above equation. We put forward the following ansatz for the density field for one soliton solution under the continuum limit which is,
\bea
\rho(x)=\rho_o(x)+\frac{1}{2i\pi L}\left[\coth\left(\frac{x-{i\lambda}}{L}\right)-\coth\left(\frac{x+{i\lambda}}{L}\right)\right]
\label{anzsol}
\eea

where $\lambda$ is some function of the position of the dual variable that can be found from Eq.~\ref{ansatzcheck} by using this ansatz. Note that, as input, we specify the value of the dual variable $z(t=0)=a+i b $. Since this analysis is witout an external trap, therefore due to translational invariance we can assume $a=0$ (without loss of generality).
Our aim is to evaluate the L.H.S of Eq.~\ref{ansatzcheck} using the ansatz for $\rho(x)$. The Hilbert transform is solved in detail in Appendix C. It can be shown that
\bea
\label{hilbert2}
\rho(x)^{\mathrm{H}}=-\frac{1}{2\pi L}\left[\coth\left(\frac{x-{i\lambda}}{L}\right)+\coth\left(\frac{x+{i\lambda}}{L}\right)\right]
\eea 
Now, the task is to solve the other part of L.H.S,
$-\frac{g}{2}\frac{1}{\rho(x)}\rho'(x)$,
Plugging in the ansatz (Eq.~\ref{anzsol}), we get, 
\bea
\frac{1}{2}\frac{\rho'(x)}{\rho(x)}=\frac{1}{\rho(x)}\left(-\frac{1}{{4i}\pi L^{2}}\right)\left[\coth^{2}\left(\frac{x-i\lambda}{L}\right)-\coth^{2}\left(\frac{x+i\lambda}{L}\right)\right]
\eea
which implies, 
\bea
 g&&\Bigg(\pi \rho(x)^{\mathrm{H}}-\frac{\partial}{\partial x}\ln\sqrt{\rho(x)}\Bigg)=\left(-\frac{g}{2L}\right)\Bigg[\coth\left(\frac{x-i\lambda}{L}\right)\nonumber\\
&&+\coth\left(\frac{x+i\lambda}{L}\right)\Bigg]
\left\{ 1-\frac{\left(\frac{1}{\mathit{2i}\pi L}\right)\left[\coth\left(\frac{x-i\lambda}{L}\right)-\coth\left(\frac{x+i\lambda}{L}\right)\right]}{\rho_{o}+\frac{1}{\mathit{2i}\pi L}\left[\coth\left(\frac{x-i\lambda}{L}\right)-\coth\left(\frac{x+i\lambda}{L}\right)\right]}\right\}
\eea
This gives us, 
\bea
\label{64}
 g\Bigg(\pi\rho(x)^{\mathrm{H}}&&-\frac{\partial}{\partial x}\ln\sqrt{\rho(x)}\Bigg)=\nonumber\\
&&\left(-\frac{g}{2L}\right)\left\{ \frac{2\sinh\left(\frac{2x}{L}\right)}{\cosh\left(\frac{2x}{L}\right)-\left[\cos\left(\frac{2\lambda}{L}\right)-\left(\frac{1}{\pi L\rho_{o}}\right)\sin\left(\frac{2\lambda}{L}\right)\right]}\right\} 
\eea
In order to satisfy Eq.~\ref{ansatzcheck}, we must have
\bea
\label{lambda -b relation}
\cos\left(\frac{2\lambda}{L}\right)-\left(\frac{1}{\pi L\rho_{o}}\right)\sin\left(\frac{2\lambda}{L}\right)=\cos\left(\frac{2b}{L}\right)=\cosh\left(\frac{2ib}{L}\right)
\eea
The above equation gives the relation between $\lambda$ and b. This transcendental equation can be solved numerically to get the value of $\lambda$ for certain value of b. If $\left( \frac{2\lambda}{L}\right)$ and $\left(\frac{2b}{L}\right)$ is quite small, we can approximately solve the above equation to get a functional form of $\lambda$
which gives (by Taylor expansion of Trigonometric  functions),
\bea
\label{small-lambda-b relation}
\lambda=\frac{1}{2\pi\rho_{o}}\left[\left(1+(2\pi b\rho_{o})^{2}\right)^{\frac{1}{2}}-1\right]\quad \mbox{for } \lambda,b \ll  L
\eea
So, from Eq.~\ref{64} we get,
\bea
g\Bigg(\pi\rho(x)^{\mathrm{H}}-\frac{\partial}{\partial x}\ln\sqrt{\rho(x)}\Bigg)=\left(-\frac{g}{2L}\right)\left\{ \frac{\sinh\left(\frac{2x}{L}\right)}{\frac{1}{2}\left[\cosh\left(\frac{2x}{L}\right)-\cosh\left(\frac{2ib}{L}\right)\right]}\right\} 
\eea
which implies, 
\bea
 g\left(\pi\rho^{\mathrm{H}}-\partial_{x}\mathrm{ln}\sqrt{\rho(x)}\right)=-\frac{g}{2L}\left[\coth\left(\frac{x-ib}{L}\right)+\coth\left(\frac{x+ib}{L}\right)\right]
\eea
Therefore, we prove that that our ansatz was correct. Here, $b$ is given and $\lambda$ is computed by the transcendental  Eq.~\ref{lambda -b relation}.

%
%

\subsection{Velocity of the soliton.}
In earlier sections, we had said that the dual variable drags the soliton with it and so the velocity of the soliton and the dual variable should be exactly the same. Earlier we provided an expression for the time period of the dual variable in the small $y$ limit. In this section, we will present an expression for the velocity of the soliton~\cite{stone}. This is the speed (denoted as $v_{soliton}$) at which the soliton travels. From continuity equation, one gets,  
\bea
\label{vsoliton}
v_{soliton}=\frac{\rho}{\rho-\rho_o}v(x)
\eea
Now, we get $v(x)$ from real part of Eq.~(\ref{(U+fieldeqn)}). Therefore,
\bea
v(x)=\frac{g}{2iL}\Bigg[\coth\left(\frac{x-ib}{L}\right)-\coth\left(\frac{x+ib}{L}\right)\Bigg]
\eea
Placing this in Eq.~(\ref{vsoliton}) we get,
\bea
v_{soliton}=\left(\frac{g}{2iL}\right)&&\left(\frac{\rho_{o}+\frac{1}{\mathit{2i}\pi L}\Bigg[\coth\left(\frac{x-i\lambda}{L}\right)-\coth\left(\frac{x+i\lambda}{L}\right)\Bigg]}{\frac{1}{\mathit{2i}\pi L}\Bigg[\coth\left(\frac{x-i\lambda}{L}\right)-\coth\left(\frac{x+i\lambda}{L}\right)\Bigg]}\right)\cdot\nonumber\\
&&\hspace{1in}\times\left[\coth\left(\frac{x-ib}{L}\right)-\coth\left(\frac{x+ib}{L}\right)\right]
\eea
\bea
=\pi g\rho_{o}\frac{\sinh\left(\frac{2ib}{L}\right)}{\sinh\left(\frac{2i\lambda}{L}\right)}\left(\frac{\cosh\left(\frac{2x}{L}\right)-\bigg[\cosh\left(\frac{2i\lambda}{L}\right)-\frac{1}{i\rho_{o}\pi L}\sinh\left(\frac{2i\lambda}{L}\right)\bigg]}{\cosh\left(\frac{2x}{L}\right)-\cosh\left(\frac{2ib}{L}\right)}\right)
\eea
Using Eq.~\ref{lambda -b relation} we get
\bea
v_{soliton}&&
=\pi g\rho_{o}\frac{\sinh\left(\frac{2ib}{L}\right)}{\sinh\left(\frac{2i\lambda}{L}\right)}\left(\frac{\cosh\left(\frac{2x}{L}\right)-\cosh\left(\frac{2ib}{L}\right)}{\cosh\left(\frac{2x}{L}\right)-\cosh\left(\frac{2ib}{L}\right)}\right)\\
\nonumber\\
&&
=\pi g\rho_{o}\frac{\sin\left(\frac{2b}{L}\right)}{\sin\left(\frac{2\lambda}{L}\right)}
\eea
For small values of $\left(\frac{\lambda}{L}\right)$ and $\left(\frac{b}{L}\right)$, using Eq.~(\ref{small-lambda-b relation}) we get, 
\bea
v_{soliton}=\left(\frac{2gb\left(\pi^{2}\rho_{o}^{2}\right)}{\left[\left(1+\left(2\pi b\rho_{o}\right)^{2}\right)^{\frac{1}{2}}-1\right]}\right) \quad \mbox{for } \lambda,b \ll  L
\eea
\section{Conclusions and Outlook} 
 \la{sec:conclusion}

To summarize, in this paper, we introduced the general form of HC model with an external confining box-like potential for which the system remains integrable. We showed that the box-like potential does not allow the particles to spread out even if their number increases. The typical length of the system scales as $\sinh^{-1}(\sqrt{N})$, hence logarithmically, as opposed to $\sqrt{N}$ in the Calogero-Moser system (rational).
\\
\\
We formulated the effective dual system for this model. These dual variables move in the complex plane. We showed, although the first order equations are coupled, the second order equations get completely decoupled which finally gives us the equation of motions for the Calogero particles.  We showed that the number of dual variables is independent of Calogero particles, i.e., the number of dual particles can be less than the number of Calogero particles.  By specifying the number of dual particles, we restrict the space of initial conditions and this gives us the soliton solutions. We show that $M$ dual variables give us the $M$ soliton solution. 
\\
\\
We analysed the density profile corresponding to the background, (i.e. when the dual variable is not present) and analysed its dependence on the system parameters. We also found analytical expressions for the density profile and showed its similarity to a trigonometric version that appears in the context of Gross-Witten-Wadia phase transition in large-N gauge theories \cite{gross1993possible,wadia1980n}. We then formulated the initial conditions of position and corresponding momenta for one, two, three and four soliton solution using the damping equation. We analysed the dynamics of the particles and showed that the soliton formed does not break down as the system evolves. Thus we could show that they were indeed soliton solutions for the HC model. We showed that the height of solitons depends on the proximity of the dual variable from the real axis, i.e., larger the value of the imaginary part of the dual variable, shorter is the soliton.  We also checked the effects of soliton collisions and quenching. We showed that quenching a parameter immediately breaks the soliton and generates ripples. 
\\

We explore the connection between the motion of dual variables in the complex plane and the motion of soliton. We showed that the dual variable essentially drags the soliton with it as it moves. We showed that the time period of one complete revolution for the two is equal. We solved the equation of motion of a single dual variable in the small $y$ limit and found an analytic form of the time period which matches very well with simulation. 
\\

Finally, we discuss the continuum limit. We formulated the Hamiltonian as a function of the density and velocity field. We formed an equivalent dual system using meromorphic fields and formulated their exact form in the continuum limit.  Even in this limit, the soliton solutions can be found. We formed the correct first order equation for one soliton solution. We made an ansatz for the analytic form of density for one soliton solutions and proved it to be correct. Comparisons were also made with brute force numerical simulations. \\

%
There are several directions for future and we state some of them here. Soliton stability analysis for these models is of interest and requires further investigation. Another question of interest would be to investigate if the solutions of the finite set of  equations for the background for finite number of particles correspond to zeros of known polynomials. This would be an interesting generalization of the Stieltjes problem \cite{shastry2001solution}. In the regime where the confined Hyperbolic model could be written in the Bogomolny (positive-definite) form, the HC model is deeply connected to a generalization of the Log-Gas \cite{forrester2010log} and such connections need \cite{CalogeroMosermodel} to be explored. One could also explore the possible relation of these hyperbolic models to random matrix theory \cite{bogomolny2009random}. 

\section{Acknowledgements}
We would like to thank A. Polychronakos, A. Abanov, P. Wiegmann, E. Bettelheim, E. Bogomolny, S. Majumdar, D. Huse  for useful discussions. M. K. gratefully acknowledges the Ramanujan Fellowship
SB/S2/RJN-114/2016 from the Science and Engineering Research Board
(SERB), Department of Science and Technology, Government of India. MK would like to acknowledge support from the project 6004-1 of the Indo-French Centre for the Promotion of Advanced Research (IFCPAR). We would like to thank the ICTS program on Integrable systems in Mathematics, Condensed Matter and Statistical Physics (Code: ICTS/integrability2018/07) for enabling valuable discussions with many participants. M. K. thanks the hospitality of the Department of Physics, Princeton University, USA where part of this work was done. M. K. thanks the hospitality of Laboratoire de Physique Th\'eorique, \'Ecole Normale Sup\'erieure, Paris where part of this work was done. 

\newpage
\section{Appendix A: Equation of Motion from dual equations}
In this section, we derive the equations of motion of the  particles moving in the real axis and for the dual variables moving in the complex plane from the set of dual equations. We assume that there are N real particles and $M(<N)$ number of dual variables. So, we start from the dual equation.   
\bea
\label{1stderi}
\dot{x_{i}}-i\frac{A}{L}\sinh\left(\frac{2x_{i}}{L}\right)=-i\frac{g}{L}\sum_{j\neq i}^{N}&&\coth\left(\frac{x_{i}-x_{j}}{L}\right)\nonumber\\
&&+i\frac{g}{L}\sum_{n=1}^{M}\coth\left(\frac{x_{i}-z_{n}}{L}\right)
\eea
\bea
\dot{z_{n}}-i\frac{A}{L}\sinh\left(\frac{2z_{n}}{L}\right)=i\frac{g}{L}\sum_{m\neq n}^{M}&&\coth\left(\frac{z_{n}-z_{m}}{L}\right)\nonumber\\
&&-i\frac{g}{L}\sum_{i=1}^{N}\coth\left(\frac{z_{n}-x_{i}}{L}\right)
\eea
Taking the second derivative of the Eq.~\ref{1stderi} we get
\bea
\ddot{x}_{i}=2i\frac{A}{L^{2}}\cosh\left(\frac{2x_{i}}{L}\right)\left(\frac{\dot{x}_{i}}{L}\right)&&+i\frac{g}{L^{2}}\sum_{j\neq i}^{N}\mathrm{csch}^{2}\left(\frac{x_{i}-x_{j}}{L}\right)\left(\frac{\dot{x}_{i}-\dot{x}_{j}}{L}\right)\nonumber\\
&&-i\frac{g}{L^{2}}\sum_{n=1}^{M}\mathrm{csch}^{2}\left(\frac{x_{i}-z_{n}}{L}\right)\left(\frac{\dot{x}_{i}-\dot{z}_{n}}{L}\right)
\eea
Then substituting $\dot{x}_i$ and $\dot{z}_n$ using the dual equations we get 
\bea
\label{cumbersome-eom}
\ddot{x}_{i}&&=2i\frac{A}{L^{2}}\cosh\left(\frac{2x_{i}}{L}\right)\Biggl\{i\frac{A}{L}\sinh\left(\frac{2x_{i}}{L}\right)-i\frac{g}{L}\sum_{j\neq i}^{N}\coth\left(\frac{x_{i}-x_{j}}{L}\right)\nonumber\\
&&+i\frac{g}{L}\sum_{n=1}^{M}\coth\left(\frac{x_{i}-z_{n}}{L}\right)\Biggr\}+i\frac{g}{L^{2}}\sum_{j\neq i}^{N}\mathrm{csch}^{2}\left(\frac{x_{i}-x_{j}}{L}\right)\nonumber\\
&&\Biggl\{ i\frac{A}{L}\left(\sinh\left(\frac{2x_{i}}{L}\right)-\sinh\left(\frac{2x_{j}}{L}\right)\right)-i\frac{g}{L}\Biggl[\sum_{a\neq i}^{N}\coth\left(\frac{x_{i}-x_{a}}{L}\right)\nonumber\\
&&-\sum_{b\neq j}^{N}\coth\left(\frac{x_{j}-x_{b}}{L}\right)\Biggl]+i\frac{g}{L}\sum_{n=1}^{M}\Biggl[\coth\left(\frac{x_{i}-z_{n}}{L}\right)\nonumber\\
&&-\coth\left(\frac{x_{j}-z_{n}}{L}\right)\Biggr]\Biggr\}\mathcal{-}i\frac{g}{L^{2}}\sum_{n=1}^{M}\mathrm{csch}^{2}\left(\frac{x_{i}-z_{n}}{L}\right)\Biggl[i\frac{A}{L}\Bigg(\sinh\left(\frac{2x_{i}}{L}\right)\nonumber\\
&&-\sinh\left(\frac{2z_{n}}{L}\right)\Bigg)-i\frac{g}{L}\Biggl\{\sum_{j\neq i}^{N}\coth\left(\frac{x_{i}-x_{j}}{L}\right)+\sum_{m\neq n}^{M}\coth\left(\frac{z_{n}-z_{m}}{L}\right)\Biggr\}\nonumber\\
&&+i\frac{g}{L}\Biggl\{\sum_{m=1}^{M}\coth\left(\frac{x_{i}-z_{m}}{L}\right)-\sum_{j=1}^{N}\coth\left(\frac{z_{n}-x_{j}}{L}\right)\Biggr\}\Biggr]
\eea
Now we will break the above equation in parts/categories to solve and extract the equations of motion from it. 
First we get the following expression, 
\bea
T_0=-\frac{2A^{2}}{L^{3}}\sinh\left(\frac{2x_{i}}{L}\right)\cosh\left(\frac{2x_{i}}{L}\right)
\eea
This term is already in a simplified form and is the first term in the equation of motion (Eq.~\ref{cumbersome-eom}). Then we simplify the following expression ,
\bea
T_1=2\cosh\left(\frac{2x_{i}}{L}\right)\sum_{j\neq i}^{N}\coth\left(\frac{x_{i}-x_{j}}{L}\right)&&-\sum_{j\neq i}^{N}\mathrm{csch}^{2}\left(\frac{x_{i}-x_{j}}{L}\right)\nonumber\\
&&\left(\sinh\left(\frac{2x_{i}}{L}\right)-\sinh\left(\frac{2x_{j}}{L}\right)\right)\nonumber \\
\eea
\bea
=\sum_{j\neq i}^{N}\frac{1}{\sinh\left(\frac{x_{i}-x_{j}}{L}\right)}\Biggl[2\cosh\left(\frac{2x_{i}}{L}\right)\cosh&&\left(\frac{x_{i}-x_{j}}{L}\right)\nonumber\\
&&-2\cosh\left(\frac{x_{i}+x_{j}}{L}\right)\Biggr]
\eea
\bea
=\sum_{j\neq i}^{N}\frac{1}{\sinh\left(\frac{x_{i}-x_{j}}{L}\right)}\left[\cosh\left(\frac{3x_{i}-x_{j}}{L}\right)-\cosh\left(\frac{x_{i}+x_{j}}{L}\right)\right]
\eea
\bea
=2\sum_{j\neq i}^{N}\sinh\left(\frac{2x_{i}}{L}\right)
\eea
Therefore, we get, 
\bea
 T_{1}=2\left(N-1\right)\sinh\left(\frac{2x_{i}}{L}\right)
\eea
Similarly we get a term like,
\bea
T_{2}=2\cosh\left(\frac{2x_{i}}{L}\right)\sum_{n=1}^{M}&&\coth\left(\frac{x_{i}-z_{n}}{L}\right)-\sum_{n=1}^{M}\mathrm{csch}^{2}\left(\frac{x_{i}-z_{n}}{L}\right)\nonumber\\
&&\left(\sinh\left(\frac{2x_{i}}{L}\right)-\sinh\left(\frac{2z_{n}}{L}\right)\right)
\eea
which yileds, 
\bea
 T_{2}=2\sum_{n=1}^{M}\sinh\left(\frac{2x_{i}}{L}\right)=2M\sinh\left(\frac{2x_{i}}{L}\right)
\eea
Thus, $T_0$, $T_1$ and $T_2$ are the contribution from external potential. Next, we simplify the expressions which have contribution from the interaction potential. Thus the next term we simplify is ,
\bea
T_3=\sum_{j\neq i}^{N}\mathrm{csch}^{2}\left(\frac{x_{i}-x_{j}}{L}\right)\Biggl[\sum_{a\neq i}^{N}\coth&&\left(\frac{x_{i}-x_{a}}{L}\right)\nonumber\\
&&-\sum_{b\neq j}^{N}\coth\left(\frac{x_{j}-x_{b}}{L}\right)\Biggr]
\eea
\bea
=\sum_{j\neq i}^{N}2\frac{\cosh\left(\frac{x_{i}-x_{j}}{L}\right)}{\sinh^{3}\left(\frac{x_{i}-x_{j}}{L}\right)}&&+\sum_{j\neq i}^{N}\sum_{k\neq i,j}^{N}\mathrm{csch}^{2}\left(\frac{x_{i}-x_{j}}{L}\right)\nonumber\\
&&\left[\coth\left(\frac{x_{i}-x_{k}}{L}\right)-\coth\left(\frac{x_{j}-x_{k}}{L}\right)\right]
\eea
\bea
=\sum_{j\neq i}^{N}2\frac{\cosh\left(\frac{x_{i}-x_{j}}{L}\right)}{\sinh^{3}\left(\frac{x_{i}-x_{j}}{L}\right)}+\sum_{j\neq i}^{N}\sum_{k\neq i,j}^{N}\mathrm{csch}^{2}&&\left(\frac{x_{i}-x_{j}}{L}\right)\nonumber\\
&&\left[\frac{\sinh\left(\frac{x_{j}-x_{k}-x_{i}-x_{k}}{L}\right)}{\sinh\left(\frac{x_{i}-x_{k}}{L}\right)\sinh\left(\frac{x_{j}-x_{k}}{L}\right)}\right]
\eea
\bea
=\sum_{j\neq i}^{N}2&&\frac{\cosh\left(\frac{x_{i}-x_{j}}{L}\right)}{\sinh^{3}\left(\frac{x_{i}-x_{j}}{L}\right)}\nonumber\\
&&+\sum_{j\neq i}^{N}\sum_{k\neq i,j}^{N}\left[\frac{-1}{\sinh\left(\frac{x_{i}-x_{j}}{L}\right)\sinh\left(\frac{x_{i}-x_{k}}{L}\right)\sinh\left(\frac{x_{j}-x_{k}}{L}\right)}\right]
\eea
Now
\bea
\sum_{j\neq i}^{N}\sum_{k\neq i,j}^{N}\left(\frac{-1}{\sinh\left(\frac{x_{i}-x_{j}}{L}\right)\sinh\left(\frac{x_{i}-x_{k}}{L}\right)\sinh\left(\frac{x_{j}-x_{k}}{L}\right)}\right)=0
\eea
This is because the individual terms in the summation cancels each other. ($\sinh$ is an odd function)
\bea
 T_{3}&&=\sum_{j\neq i}^{N}\mathrm{csch}^{2}\left(\frac{x_{i}-x_{j}}{L}\right)\Biggl[\sum_{a\neq i}^{N}\coth\left(\frac{x_{i}-x_{a}}{L}\right)-\sum_{b\neq j}^{N}\coth\left(\frac{x_{j}-x_{b}}{L}\right)\Biggr]\nonumber\\
&&=\sum_{j\neq i}^{N}2\frac{\cosh\left(\frac{x_{i}-x_{j}}{L}\right)}{\sinh^{3}\left(\frac{x_{i}-x_{j}}{L}\right)}
\eea
The next term is,
\bea
T_{4}&&=-\sum_{j\neq i}^{N}\mathrm{csch}^{2}\left(\frac{x_{i}-x_{j}}{L}\right)\sum_{n=1}^{M}\left[\coth\left(\frac{x_{i}-z_{n}}{L}\right)-\coth\left(\frac{x_{i}-z_{n}}{L}\right)\right]\nonumber\\
&&+\sum_{n=1}^{M}\mathrm{csch}^{2}\left(\frac{x_{i}-z_{n}}{L}\right)\sum_{j\neq i}^{N}\left[\coth\left(\frac{z_{n}-x_{j}}{L}\right)-\coth\left(\frac{x_{i}-x_{j}}{L}\right)\right]
\eea
\bea
=\sum_{j\neq i}^{N}\sum_{n=1}^{M}\Biggl[-\mathrm{csch}^{2}&&\left(\frac{x_{i}-x_{j}}{L}\right)\frac{\sinh\left(\frac{x_{j}-z_{n}-x_{i}+z_{n}}{L}\right)}{\sinh\left(\frac{x_{i}-z_{n}}{L}\right)\sinh\left(\frac{x_{j}-z_{n}}{L}\right)}\nonumber\\
&&+\mathrm{csch}^{2}\left(\frac{x_{i}-z_{n}}{L}\right)\frac{\sinh\left(\frac{x_{i}-x_{j}-z_{n}+x_{j}}{L}\right)}{\sinh\left(\frac{z_{n}-x_{j}}{L}\right)\sinh\left(\frac{x_{i}-x_{j}}{L}\right)}\Biggr]
\eea
\bea
=\sum_{j\neq i}^{N}\sum_{n=1}^{M}\Biggl[&&\frac{1}{\sinh\left(\frac{x_{i}-x_{j}}{L}\right)\sinh\left(\frac{x_{i}-z_{n}}{L}\right)\sinh\left(\frac{x_{j}-z_{n}}{L}\right)}\nonumber\\
&&-\frac{1}{\sinh\left(\frac{x_{i}-x_{j}}{L}\right)\sinh\left(\frac{x_{i}-z_{n}}{L}\right)\sinh\left(\frac{x_{j}-z_{n}}{L}\right)}\Biggr]=0
\eea
At last we simplify the following term,
\bea
T_{5}=\sum_{n=1}^{M}\sum_{m\neq n}^{M}\Biggl[-\mathrm{csch}^{2}\left(\frac{x_{i}-z_{n}}{L}\right)&&\Biggl\{ \coth\left(\frac{z_{n}-z_{m}}{L}\right)\nonumber\\
&&-\coth\left(\frac{x_{i}-z_{m}}{L}\right)\Biggr\} \Biggr]
\eea
\bea
=\sum_{n=1}^{M}\sum_{m\neq n}^{M}\Biggl[-\mathrm{csch}^{2}\left(\frac{x_{i}-z_{n}}{L}\right)\left\{ \frac{\sinh\left(\frac{x_{i}-z_{m}-z_{n}+z_{m}}{L}\right)}{\sinh\left(\frac{z_{n}-z_{m}}{L}\right)\sinh\left(\frac{x_{i}-z_{n}}{L}\right)}\right\} \Biggr]
\eea
which gives us, 
\bea
 T_{5}=\sum_{n=1}^{M}\sum_{m\neq n}^{M}\left[\frac{-1}{\sinh\left(\frac{x_{i}-z_{n}}{L}\right)\sinh\left(\frac{z_{n}-z_{m}}{L}\right)\sinh\left(\frac{x_{i}-z_{n}}{L}\right)}\right]=0
\eea
Now we have all the simplified terms of the Eq.~\ref{cumbersome-eom}. Summing $T_0$ to $T_5$ we get,
\bea
\ddot{x_{i}}&&=-\frac{2A^{2}}{L^{3}}\sinh\left(\frac{2x_{i}}{L}\right)\cosh\left(\frac{2x_{i}}{L}\right)+\frac{2Ag}{L^{3}}(N-M-1)\sinh\left(\frac{2x_{i}}{L}\right)\nonumber\\
&&+\frac{2g^{2}}{L^{3}}\sum_{j\neq i}^{N}\left(\frac{\cosh\left(\frac{x_{i}-x_{j}}{L}\right)}{\sinh^{3}\left(\frac{x_{i}-x_{j}}{L}\right)}\right)
\eea
This entire process can be applied for finding the equation of motion for the dual variables. There will be a change from the contribution of external potential as the corresponding summation limits will change as there are M dual variables instead of N. Nevertheless the forms are quite similar and after all the algebra we get,
\bea
\ddot{z_{n}}&&=-\frac{2A^{2}}{L^{3}}\sinh\left(\frac{2z_{n}}{L}\right)\cosh\left(\frac{2z_{n}}{L}\right)+\frac{2Ag}{L^{3}}(N-M+1)\sinh\left(\frac{2z_{n}}{L}\right)\nonumber\\
&&+\frac{2g^{2}}{L^{3}}\sum_{m\neq n}^{M}\left(\frac{\cosh\left(\frac{z_{n}-z_{m}}{L}\right)}{\sinh^{3}\left(\frac{z_{n}-z_{m}}{L}\right)}\right)
\eea
\section{Appendix B: Field Theory}
\subsection{Part 1: Formation of $U^+(x)$ in the continuum limit.} 
For large number of particles we can define a smooth continuous position function $x(s)$ such that $x(j)=x_j$. For $N\rightarrow\infty$ the position function becomes unique \cite{polymanas17} and is related to the density of the system as, 
\bea
x'(j)=\frac{dx}{dj}=\frac{1}{\rho(x)}
\eea

We start from Eq.~\ref{HlimitU(x)},
\bea
\label{hysum1}
\rho(x)U^{+}(x)=\rho(x)v(x)+\frac{ig}{L}\lim_{N\rightarrow\infty}\sum_{  k\neq j\atop k=-N}^{N}\rho(x)\coth\left(\frac{x(j)-x(k)}{L}\right)
\eea
The function $\coth \big( \frac{x-a }{L}\big)$ has a simple pole at $x=a$. We can see this in the Laurent  expansion of $\coth  \big( \frac{x-a}{L}  \big)$.
\bea
\coth\left(\frac{x-x_{k}}{L}\right)=\frac{1}{\left(\frac{x-x_{k}}{L}\right)}+\frac{\left(\frac{x-x_{k}}{L}\right)}{3}-\frac{\left(\frac{x-x_{k}}{L}\right)^{3}}{45}+\mathcal{O}\left[x^{5}\right] 
\eea
Therefore, we can write the Eq.~\ref{hysum1} as,
\bea
\label{hysum2}
\lim_{N\rightarrow\infty}\sum_{k\neq j\atop k=-N}^{N}\coth\left(\frac{x(j)-x(k)}{L}\right)=\sum_{k=-\infty}^{\infty}f(k)-\lim_{k\rightarrow j}f(k)
\eea
where $f(k)$ is
\bea
\label{f(k)}
f(k)=\coth\left(\frac{x(j)-x(k)}{L}\right)+\frac{L}{x'(j)(k-j)}
\eea
It is important to note that the above treatment is only valid for $N\rightarrow\infty$ because, only then, for any $j$, we will have, 
\bea
\sum_{k=-\infty}^{\infty}\frac{L}{x'(j)(k-j)}=0
\eea
The function $f(k)$ is so chosen such that the limit at $k\rightarrow j$ exists. The limiting value can be found out in the following way,
\bea
\lim_{k\rightarrow j}f(k)=\lim_{k\rightarrow j}\Biggl[\frac{1}{\left(\frac{x(j)-x(k)}{L}\right)}&+&\frac{\left(\frac{x(j)-x(k)}{L}\right)}{3}-\frac{\left(\frac{x(j)-x(k)}{L}\right)^{3}}{45}\nonumber\\
&+&\mathcal{O}\left[x^{5}\right]+\frac{L}{x'(j)(k-j)}\Biggr]
\eea
The only non-trivial terms remaining are,
\bea
\lim_{k\rightarrow j}f(k)=\lim_{k\rightarrow j}\left[\frac{L}{\left(x(j)-x(k)\right)}+\frac{L}{x'(j)(k-j)}\right]
\eea
From the Taylor expansion of $x(k)$ we get, 
\bea
\lim_{k\rightarrow j}f(k)=L\frac{x''(j)}{2\left[x'(j)\right]^{2}}
\eea
Therefore from Eq.~\ref{hysum2}
\bea
\lim_{N\rightarrow\infty}\sum_{k\neq j\atop k=-N}^{N}\coth\left(\frac{x(j)-x(k)}{L}\right)=\int_{-\infty}^{\infty}f(\nu)d\nu-L\frac{x''(j)}{2\left[x'(j)\right]^{2}}
\eea
Now we will solve the integral in the above equation.
\bea
\int_{-\infty}^{\infty}f(\nu)d\nu=\int_{-\infty}^{j-\epsilon}f(\nu)d\nu+\int_{j+\epsilon}^{\infty}f(\nu)d\nu+\lim_{\epsilon\rightarrow0}\int_{j-\epsilon}^{j+\epsilon}f(j)d\nu
\eea
As the limiting value of $f(k)$ as $k\rightarrow j$ is finite we have,
\bea
\lim_{\epsilon\rightarrow0}\int_{j-\epsilon}^{j+\epsilon}f(j)d\nu=0
\eea
Therefore we can convert the above integral into a principal value integral. By doing so, we can replace $f(k)$ by $\coth\left(\frac{x(j)-x(\nu)}{L}\right)$ as the other part contributes $0$ to the principal value integral.
\bea
\int_{-\infty}^{j-\epsilon}f(\nu)d\nu+\int_{j+\epsilon}^{\infty}f(\nu)d\nu&&=P\left\{ \int_{-\infty}^{\infty}f(\nu)d\nu\right\}\nonumber\\
&& =P\left\{ \int_{-\infty}^{\infty}\coth\left(\frac{x(j)-x(\nu)}{L}\right)d\nu\right\} 
\eea
Let $x(\nu)=\tau$ .Therefore $d\nu=\rho(\tau)d\tau$. Hence
\bea
P\left\{ \int_{-\infty}^{\infty}\coth\left(\frac{x(j)-x(\nu)}{L}\right)d\nu\right\} &&=P\left\{ \int_{-\infty}^{\infty}\coth\left(\frac{x-\tau}{L}\right)\right\} \rho(\tau)d\tau\nonumber\\
&&=-\pi L\rho(x)^{\mathrm{H}}
\eea
The next task is to find the limiting value in terms of the density field. We know $x'(j)=\frac{1}{\rho(x)}$. Therefore
\bea
x''(j)
=-\frac{\rho'(x)}{\left[\rho(x)\right]^{3}}
\eea
This implies,
\bea
L\frac{x''(j)}{2\left[x'(j)\right]^{2}}
=-L\frac{\partial}{\partial x}\ln\sqrt{\rho(x)}
\eea
Therefore we finally get
\bea
\lim_{N\rightarrow\infty}\sum_{k\neq j\atop k=-N}^{N}\coth\left(\frac{x(j)-x(k)}{L}\right)=L\frac{\partial}{\partial x}\ln\sqrt{\rho(x)}-\pi L\rho(x)^{\mathrm{H}}
\eea
Hence from Eq.~\ref{hysum1} we get
\bea
\rho(x)U^{+}(x)=\rho(x)v(x)+i\frac{g}{L}\rho(x)\left(L\frac{\partial}{\partial x}\ln\sqrt{\rho(x)}-\pi L\rho(x)^{\mathrm{H}}\right)
\eea
\subsection{Part 2: Formulation of the Hamiltonian}
We begin with the interaction potential in the Hamiltonian for limited number of particles. From that we will find the equivalent form under the continuum limit using density fields. The interaction potential as we know is of the form
\bea
V_{int}=\sum_{j=1}^{N}\sum_{k\neq j\atop k=1}^{N}\frac{g^{2}}{2L^{2}}\left(\frac{1}{\sinh^{2}\left(\frac{x_j-x_k}{L}\right)}\right)
\eea
Under continuum limit the positions gets replaced by position functions as before and also the summation ranges from $-\infty$ to $\infty$ as $N\rightarrow\infty$. Hence, $V_{int}$ becomes
\bea
\label{Actualvint}
V_{int}=\lim_{N\rightarrow\infty}\sum_{j=-N}^{N}\sum_{k\neq j\atop k=-N}^{N}\frac{g^{2}}{2L^{2}}\left(\frac{1}{\sinh^{2}\left(\frac{x(j)-x(k)}{L}\right)}\right)
\eea
We first attempt to find the result of the following summation 
\bea
\lim_{N\rightarrow\infty}\sum_{k\neq j\atop k=-N}^{N}\frac{g^{2}}{2L^{2}}\left(\frac{1}{\sinh^{2}\left(\frac{x(j)-x(k)}{L}\right)}\right)\nonumber
\eea 
From the Taylor expansion of $\left(\frac{1}{\sinh^{2}\left(\frac{x(j)-x(k)}{L}\right)}\right)$, we can show that it has a pole of second order,
\bea
\frac{1}{\sinh^{2}\left(\frac{x(j)-x(k)}{L}\right)}=\frac{1}{\left(\frac{x(j)-x(k)}{L}\right)^{2}}&&-\frac{1}{3}+\frac{\left(\frac{x(j)-x(k)}{L}\right)^{2}}{15}\nonumber\\
&&+\mathcal{O}\left[\left(\frac{x(j)-x(k)}{L}\right)^{4}\right]
\eea
As before we define a different function $g(k)$ which can be related to our required sum. The function $g(k)$ is not singular and hence is easier to work with. We define g(k) as,
\bea
g(k)=\frac{1}{\sinh^{2}\left(\frac{x(j)-x(k)}{L}\right)}-\frac{L^{2}}{\left[x'(j)\right]^{2}(k-j)^{2}}+\frac{x''(j)}{\left[x'(j)\right]^{3}}\frac{L^{2}}{(k-j)}
\eea
So, we can write,
\bea
\label{vint}
\lim_{N\rightarrow\infty}\sum_{k\neq j\atop k=-N}^{N}\left(\frac{1}{\sinh^{2}\left(\frac{x(j)-x(k)}{L}\right)}\right)=&&\sum_{k=-\infty}^{\infty}g(k)\nonumber\\
&&-\lim_{k\rightarrow j}g(k)+\frac{\pi^{2}}{3}\frac{L^{2}}{\left[x'(j)\right]^{2}}
\eea
It is important to note the origin of the last term of the above equation. Unlike Eq.~\ref{f(k)}, here in $g(k)$ there is an even term as well. The odd term still contributes $0$ to the summation but the even term is not equal to $0$ when we sum it over from $\infty$ to $-\infty$. Instead we have
\bea
\sum_{k=-\infty}^{\infty}\frac{L^{2}}{\left[x'(j)\right]^{2}(k-j)^{2}}=\frac{L^{2}}{\left[x'(j)\right]^{2}}\sum_{k=-\infty}^{\infty}\frac{1}{(k-j)^2}
\eea 
Replacing $(k-j)=s$ and since the summation is from $\infty$ to $-\infty$, we can write the above summation as \cite{stone},
\bea
\frac{L^{2}}{\left[x'(j)\right]^{2}}\sum_{k=-\infty}^{\infty}\frac{1}{(k-j)^2}=\frac{L^{2}}{\left[x'(j)\right]^{2}}\sum_{s=-\infty}^{\infty}\frac{1}{s^2}&&
=\frac{\pi^{2}}{3}\frac{L^{2}}{\left[x'(j)\right]^{2}}
\eea
So it was required to add this term in Eq.~\ref{vint} so as that we can replace $g(k)$ properly.
We will now find the limiting value of the function $g(k)$ as $k\rightarrow j$ in the same way as was done in Part 1 above.
\bea
\lim_{k\rightarrow j}&&g(k)=\lim_{k\rightarrow j}\Biggl[\frac{1}{\left(\frac{x(j)-x(k)}{L}\right)^{2}}-\frac{1}{3}+\frac{\left(\frac{x(j)-x(k)}{L}\right)^{2}}{15}\nonumber\\
&&+\mathcal{O}\left[\left(\frac{x(j)-x(k)}{L}\right)^{4}\right]-\frac{L^{2}}{\left[x'(j)\right]^{2}(k-j)^{2}}+\frac{x''(j)}{\left[x'(j)\right]^{3}}\frac{L^{2}}{(k-j)}\Biggr]
\eea
The non-trivial terms left are,
\bea
\lim_{k\rightarrow j}g(k)=\lim_{k\rightarrow j}\Biggl[\frac{1}{\left(\frac{x(j)-x(k)}{L}\right)^{2}}&&-\frac{L^{2}}{\left[x'(j)\right]^{2}(k-j)^{2}}\nonumber\\
&&+\frac{x''(j)}{\left[x'(j)\right]^{3}}\frac{L^{2}}{(k-j)}\Biggr]
\eea
After doing some algebra we get,
\bea
\label{limitgx}
\lim_{k\rightarrow j}g(k)&&=L^{2}\left[\frac{3}{4}\frac{\left[x''(j)\right]^{2}}{\left[x'(j)\right]^{4}}-\frac{1}{3}\frac{x'''(j)}{\left[x'(j)\right]^{3}}\right]\nonumber\\
&&=L^{2}\left[-\frac{1}{4}\frac{\left[x''(j)\right]^{2}}{\left[x'(j)\right]^{4}}-\frac{1}{3}\frac{d}{ds}\left(\frac{x''(s)}{\left[x'(s)\right]^{3}}\right)\Bigg|_{s=j}\right]
\eea
Now we need to compute $\sum_{k=-\infty}^{\infty}g(k)$. We follow the same convention as was used in the calculation of $\sum_{k=-\infty}^{\infty}f(k)$ in part 1. Here also we can exclude the part of $g(k)$ which makes it finite as $k\rightarrow j$ provided we take the principal value of the integral.
\bea
\sum_{k=-\infty}^{\infty}g(k)&&=-\frac{\partial}{\partial x}P\left\{ \int_{-\infty}^{\infty}\coth\left(\frac{x(j)-\tau}{L}\right)\rho(\tau)d\tau\right\}
\eea
Therefore, we have, 
\bea
\label{integralgx}
 \sum_{k=-\infty}^{\infty}g(k)=\pi L\frac{\partial}{\partial x}\left[\rho(x(j))^{\mathrm{H}}\right]
\eea
After finding this we go back to Eq.~\ref{Actualvint}.
\bea
\lim_{N\rightarrow\infty}\sum_{j=-N}^{N}\sum_{k\neq j\atop k=-N}^{N}&&\left(\frac{1}{\sinh^{2}\left(\frac{x(j)-x(k)}{L}\right)}\right)\nonumber\\
&&=\int_{-\infty}^{\infty}dj\sum_{k\neq j\atop k=-N}^{N}\left(\frac{1}{\sinh^{2}\left(\frac{x(j)-x(k)}{L}\right)}\right)
\eea
Using Eq.~\ref{vint}, Eq.~\ref{integralgx} and Eq.~\ref{limitgx} we have, 
\bea
I&&=\lim_{N\rightarrow\infty}\sum_{j=-N}^{N}\sum_{k\neq j\atop k=-N}^{N}\left(\frac{1}{\sinh^{2}\left(\frac{x(j)-x(k)}{L}\right)}\right)\nonumber\\
&&=\int_{-\infty}^{\infty}\Bigg[\pi L^{2}\frac{\partial}{\partial x}\left[\rho\bigg(x(j)\bigg){}^{\mathrm{H}}\right]
-L^{2}\Bigg\{-\frac{1}{4}\frac{\left[\rho'(x(j))\right]^{2}}{\left[\rho(x(j))\right]^{2}}\nonumber\\
&&\hspace{1.5in}-\frac{1}{3}\frac{d}{ds}\left(\frac{x''(s)}{\left[x'(s)\right]^{3}}\right)\Bigg|_{s=j}\Bigg\}
+L^{2}\frac{\pi^{2}}{3}\rho^{2}\bigg(x(j)\bigg)\Bigg]dj
\eea
Now from $x'(j)=\frac{1}{\rho(x)}$, we have $dj=\rho(x)dx$. Therefore,
\bea
 I=\int_{-\infty}^{\infty}\Biggl[\pi L^{2}\frac{\partial}{\partial x}\left[\rho(x){}^{\mathrm{H}}\right]+L^{2}\frac{1}{4}\frac{\left[\rho'(x)\right]^{2}}{\left[\rho(x)\right]^{2}}+L^{2}\frac{\pi^{2}}{3}\left[\rho(x)\right]^{2}\Biggr]\rho(x)dx
\eea
\bea
\tab=\int_{-\infty}^{\infty}\Biggl[\pi\frac{g^{2}}{2}\rho(x)\frac{\partial}{\partial x}\left[\rho(x){}^{\mathrm{H}}\right]+\frac{g^{2}}{8}\frac{\left[\rho'(x)\right]^{2}}{\rho(x)}+\frac{\pi^{2}g^{2}}{6}\left[\rho(x)\right]^{3}\Biggr]dx
\eea
Therefore the Hamiltonian under the assumption of continuum limit becomes,
\bea
{\cal H}=\int_{-\infty}^{\infty}dx\left[\frac{1}{2}\rho v^{2}+\frac{\pi g^{2}}{2}\rho\frac{\partial}{\partial x}\rho{}^{\mathrm{H}}+\frac{g^{2}}{8}\frac{\left[\rho'\right]^{2}}{\rho}+\frac{\pi^{2}g^{2}}{6}\rho^{3}\right]
\eea
The $\frac{1}{2}\rho v^{2}$ term comes from the kinetic energy part which is self-evident.\\
The above equation can be equivalently written in the form \cite{abanov09},
\bea
{\cal H}=\int dx\rho(x)\left[\frac{v^{2}}{2}+\frac{1}{2}\left(\pi g\rho^{H}-g\partial_{x}log\sqrt{\rho(x)}\right)^{2}\right]+const
\eea
\section{Appendix C: Hilbert transform}
\begin{figure} 
	\centering
	\includegraphics[width=0.8\linewidth]{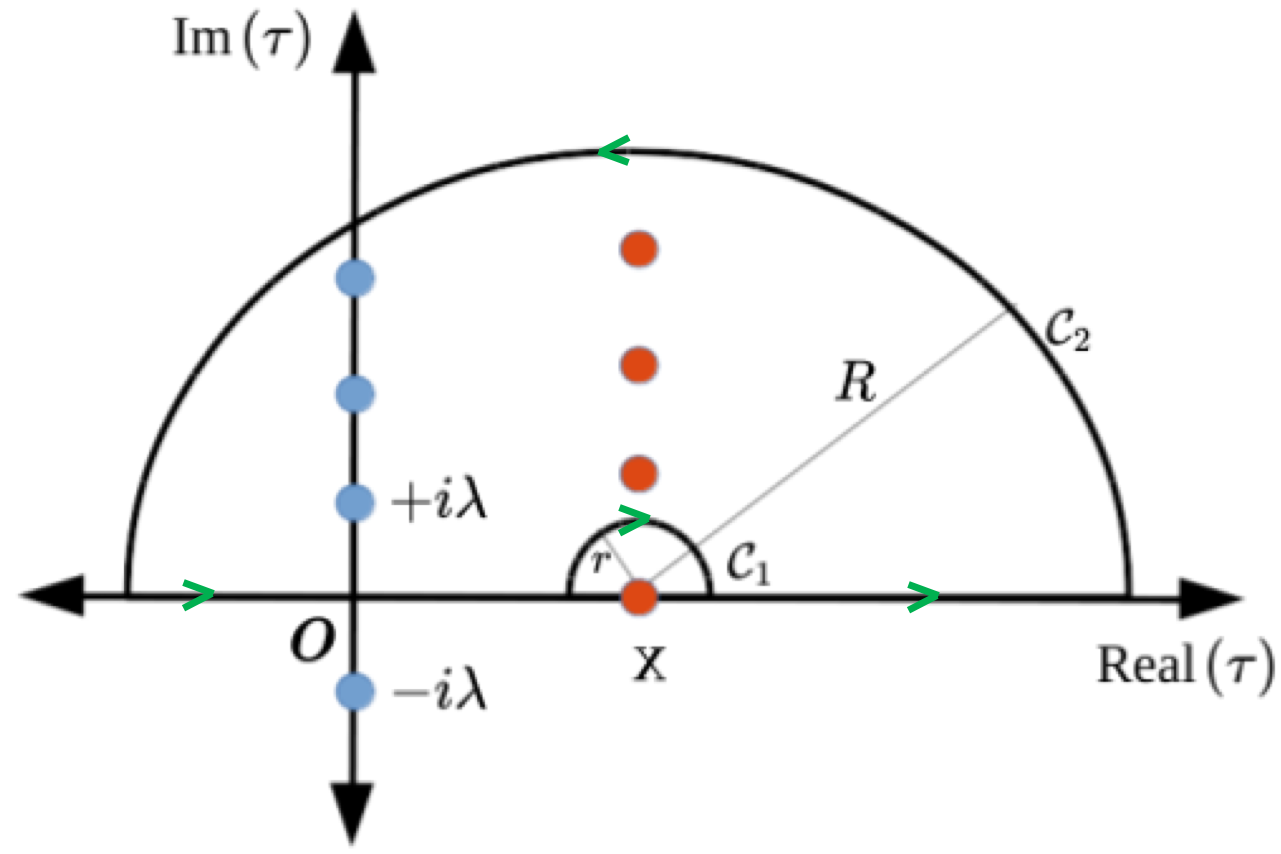}
	\caption{The contour used for solving the Hilbert transform. The contour is in the $\tau$ plane. We performed the integration going in the anticlockwise direction. Blue points denote the poles $\tau=i(\lambda+nL\pi)$ and red points denote poles $\tau=x+inL\pi,$$\;\;$$n\epsilon I^+$. Note that the pole $-i\lambda$ lies outside the chosen contour.}
	\label{fig:contour_plot1}
\end{figure}

We will first explain the Hilbert transform in general using a kernel $K\left(\frac{\tau-x}{L}\right)$ which is singular along the real axis. We use a general function $f(z)$ whose singular points are not in the real axis and the poles are simple. Therefore we define the Hilbert transform of the function $f(z)$ as,
\bea
f(x)^{\mathrm{H}}=\frac{1}{\pi L}P\left\{ \int_{-\infty}^{\infty}f(\tau)K\left(\frac{\tau-x}{L}\right)d\tau\right\} 
\eea 
where $P$ stands for the principal value of the integral. Since the integrand has poles along the real line the integral is valid only in the principal value sense. The principal value of the integral is defined as,
\bea
P\left\{ \int_{-\infty}^{\infty}f(\tau)K\left(\frac{\tau-x}{L}\right)d\tau\right\} =\lim_{\epsilon\rightarrow0}&&\Biggl[\int_{-\infty}^{x-\epsilon}f(\tau)K\left(\frac{\tau-x}{L}\right)d\tau\nonumber\\
&&+\int_{x+\epsilon}^{\infty}f(\tau)K\left(\frac{\tau-x}{L}\right)d\tau\Biggr]
\eea
Now let $g(\tau)=f(\tau)K\left(\frac{\tau-x}{L}\right)$. Therefore, 
\bea
\int_{-\infty}^{\infty}g(\tau)d\tau=P\left\{ \int_{-\infty}^{\infty}g(\tau)d\tau\right\} +\lim_{r\rightarrow0}\int_{c_{1}}g(\tau)d\tau
\eea
where $r$ is the radius of the contour $c_1$. Now since $g(\tau)$ has a simple pole at $\tau=x$, the Laurent series expansion of $g(\tau)$ is,
\bea
g(\tau)=\frac{a_{-1}}{(\tau-x)}+a_{0}+a_{1}(\tau-x)+\cdots\cdots
\eea
We will first solve the integral along the $c_1$ contour. Let $(\tau-x)=re^{i\theta}$. Therefore $d\tau=ire^{i\theta}d\theta$. \\
\bea
 \lim_{r\rightarrow0}\int_{c_{1}}g(\tau)d\tau=\lim_{r\rightarrow0}\int_{\pi}^{0}d\theta ire^{i \theta}\left[\frac{a_{-1}}{re^{i\theta}}+a_{0}+a_{1}re^{i\theta}+.......\right]
\eea
\bea
\label{I_over}
 \lim_{r\rightarrow0}\int_{c_{1}}g(\tau)d\tau=ia_{-1}\int_{\pi}^{0}d\theta=-i\pi a_{-1}
\eea
We use  residue theorem to calculate the principal value. To use the residue theorem, we need a closed contour. But, the Hilbert transform is defined on the real axis from $-\infty$ to $\infty$. Therefore we need to choose a closed contour in which integral from other parts of the contour will be equal to $0$. We choose such a contour in the upper half plane as shown in the Fig.~\ref{fig:contour_plot1}. The radius of contour $c_2$ is $R$. We evaluate the integral in limit $R\rightarrow\infty$.
Since $g(\tau)$ decays faster than than $\tau$ as $R\rightarrow\infty$ we have,
\bea
\lim_{R\rightarrow\infty}\int_{c_{2}}g(\tau)d\tau=0
\eea 
We have also checked this to be true via a rigorous evaluation of the integral for our case. 
Therefore this allows us to write the principal value integral as a contour integral over a closed contour.
\bea
\lim_{\epsilon\rightarrow 0}\int_{-\infty}^{x-\epsilon}g(\tau)d\tau+\lim_{\epsilon\rightarrow 0}\int_{x+\epsilon}^{\infty}&&g(\tau)d\tau+\lim_{r\rightarrow0}\int_{c_{1}}g(\tau)d\tau\nonumber\\
&&+\lim_{R\rightarrow\infty}\int_{c_{2}}g(\tau)d\tau=\oint g(\tau)d\tau
\eea
\bea
\label{PV_int}
\lim_{\epsilon\rightarrow 0}\int_{-\infty}^{x-\epsilon}g(\tau)d\tau+\lim_{\epsilon\rightarrow 0}&&\int_{x+\epsilon}^{\infty}g(\tau)d\tau\nonumber\\
&&=\oint g(\tau)d\tau-\lim_{r\rightarrow0}\int_{c_{1}}g(\tau)d\tau
\eea
Now from residue theorem,
\bea
\label{residue_1}
\oint g(\tau)d\tau=2\pi
i\,\sum_{i}Res\left[g(\tau),z_{i}\right]
\eea 
where $z_i's$ are the poles of $g(\tau)$ inside the closed contour, $\mathrm{Im}\left(z_{i}\right)\neq0$. Therefore using Eq.~\ref{I_over}, Eq.~\ref{PV_int} and Eq.~\ref{residue_1} we get
\bea
\lim_{\epsilon\rightarrow 0}\int_{-\infty}^{x-\epsilon}g(\tau)d\tau+\lim_{\epsilon\rightarrow 0}&&\int_{x+\epsilon}^{\infty}g(\tau)d\tau=\nonumber\\
&&2\pi
i\,\sum_{i}Res\left[g(\tau),z_{i}\right]+i\pi a_{-1}
\eea
\bea
\label{contour_final}
P\left\{ \int_{-\infty}^{\infty}g(\tau)d\tau\right\}=2\pi
i\,\sum_{i}Res\left[g(\tau),z_{i}\right]+i\pi a_{-1}
\eea
In the present case we have ,
\bea
\label{hilbert1}
\rho(x)^{\mathrm{H}}=\frac{1}{\pi L}P\left\{ \int_{-\infty}^{\infty}\rho(\tau)\coth\left(\frac{\tau-x}{L}\right)d\tau\right\}  
\eea 
and
\bea
\rho(x)=\rho_o+\frac{1}{2i\pi L}\left[\coth\left(\frac{x-i\lambda}{L}\right)-\coth\left(\frac{x+i\lambda}{L}\right)\right] 
\eea
Now the Hilbert transform of $\rho_o$ is,
\bea
\rho_{o}^{\mathrm{H}}=\frac{1}{\pi L}P\left\{ \int_{-\infty}^{\infty}\rho_{o}\coth\left(\frac{\tau-x}{L}\right)d\tau\right\} 
\eea
Now changing the variable $\left(\frac{\tau-x}{L}\right)=z$ we get,
\bea
\rho_{o}^{\mathrm{H}}=\frac{1}{\pi}P\left\{ \int_{-\infty}^{\infty}\rho_{o}\coth(z)dz\right\}=0
\eea
as $\coth (z)$ is an odd function of z. Now the function $\coth\left(\frac{x-a}{L}\right)$ has a simple pole at $x=a$, i.e. on the real axis. We can see this in the Laurent  expansion of $\coth \big(\frac{x-a}{L} \big)$.
\bea
\coth\left(\frac{x-x_{k}}{L}\right)=\frac{1}{\left(\frac{x-x_{k}}{L}\right)}+\frac{\left(\frac{x-x_{k}}{L}\right)}{3}-\frac{\left(\frac{x-x_{k}}{L}\right)^{3}}{45}+\mathcal{O}\left[x^{5}\right] 
\eea
The function $\coth\left(\frac{x-x_{k}}{L}\right)$ also has simple poles inside contour at $x=x_{k}+inL\pi$, where $(n\;\epsilon\;I^+)$. From Eq.~\ref{hilbert1} we get
\bea
\rho(x)^{\mathrm{H}}=\frac{1}{2i\pi^{2}L^{2}}&&\Biggl[P\left\{ \int_{-\infty}^{\infty}\coth\left(\frac{\tau-i\lambda}{L}\right)\coth\left(\frac{\tau-x}{L}\right)d\tau\right\}\nonumber\\
&& -P\left\{ \int_{-\infty}^{\infty}\coth\left(\frac{\tau+i\lambda}{L}\right)\coth\left(\frac{\tau-x}{L}\right)d\tau\right\}\Biggr]
\eea
Now, we have two separate Hilbert transform terms. For the first one the poles are at $\tau=x$ and $\tau=i\lambda$. For the second transform the poles are at $\tau=x$ and $\tau=-i\lambda$. The point $(-i\lambda)$ lies outside the contour. Hence its contribution to the residue will be 0. For the two separate integral we have.
\bea
a_{-1}=L\coth\left(\frac{x-i\lambda}{L}\right) \textrm{\hspace{0.5in}and\hspace{0.5in}} a_{-1}=L\coth\left(\frac{x+i\lambda}{L}\right) 
\eea
respectively\\
Thus using equation Eq.~\ref{contour_final} we get
\bea
\rho(x)^{\mathrm{H}}=\frac{1}{2i\pi^{2}L^{2}}&&\Biggl[\left\{ i\pi L\coth\left(\frac{x-i\lambda}{L}\right)-2i\pi L\coth\left(\frac{x-i\lambda}{L}\right)\right\}\nonumber\\
&& -\left\{ i\pi L\coth\left(\frac{x+i\lambda}{L}\right)\right\}\Biggr]
\eea 
\bea
\rho(x)^{\mathrm{H}}=-\frac{1}{2\pi L}\left[\coth\left(\frac{x-i\lambda}{L}\right)+\coth\left(\frac{x+i\lambda}{L}\right)\right]
\label{rhresult}
\eea 
Now for poles at $\tau=i(\lambda+nL\pi)$ and at $\tau=x+inL\pi$ (since these are Hyperbolic functions), for each n, we have the following situaton. For the term $\coth\left(\frac{\tau-i\lambda}{L}\right)\coth\left(\frac{\tau-x}{L}\right)$ the contribution from the residue is proportial to,
\bea
\coth\left(\frac{x-i\lambda+iLn\pi}{L}\right)+\coth\left(\frac{i\lambda-x+inL\pi}{L}\right)=0
\eea
Similarly, for the term $\coth\left(\frac{\tau+i\lambda}{L}\right)\coth\left(\frac{\tau-x}{L}\right)$, the contribution of the residue is equal to 0. In our case, one could also perform the Hilbert tranform as a real line integral using the basic definition of principle value to arrive at the same Eq.~\ref{rhresult}.

\section{Appendix D: Numerical techniques}
For our HC Model we have an second order differential equation- Eq.~\ref{xddot} which is of the form,
\bea
\ddot{x_i}=f(x_i,x_j)
\eea
where $i$ denotes the $i^{th}$ particle.
There is no explicit time dependence in our differential equation. In order to perform RK4  \cite{jensen} we have to break the above equation into two sets of first order differential equations which are equivalent to the Hamilton's equation of motion. Thus we have,
\bea
\dot{x_i}=p_i\;\textrm{and}\;\dot{p_i}=f(x_i,x_j)
\eea 
where 
\bea
f(x_i,x_j)&&=-\frac{2A^{2}}{L^{3}}\sinh\left(\frac{2x_{i}}{L}\right)\cosh\left(\frac{2x_{i}}{L}\right)+\nonumber\\
&&\frac{2Ag}{L^{3}}(N-M-1)\sinh\left(\frac{2x_{i}}{L}\right)+\frac{2g^{2}}{L^{3}}\sum_{j\neq i}^{N}\left(\frac{\cosh\left(\frac{x_{i}-x_{j}}{L}\right)}{\sinh^{3}\left(\frac{x_{i}-x_{j}}{L}\right)}\right)
\eea
These equation are coupled to each other. So, here the RK4 steps for each particle are as follows,
\bea
k_1(i)=dt\cdot(p_i)\hspace{1.25in}
q_1(i)=dt\cdot f(x_i,x_j)\nonumber\\
k_2(i)=dt\cdot(p_i+0.5\cdot q_1(i))\hspace{0.4in}
q_2(i)=dt\cdot\left(f(x_i,x_j)+0.5\cdot k_1(i)\cdot\frac{\mathrm{d}f}{\mathrm{d}x_i}\right) \nonumber\\
k_3(i)=dt\cdot(p_i+0.5\cdot q_2(i))\hspace{0.4in}
q_3(i)=dt\cdot\left(f(x_i,x_j)+0.5\cdot k_2(i)\cdot\frac{\mathrm{d}f}{\mathrm{d}x_i}\right) \nonumber\\
k_4(i)=dt\cdot(p_i+q_3(i))\hspace{0.72in}
q_4(i)=dt\cdot\left(f(x_i,x_j)+k_3(i)\cdot\frac{\mathrm{d}f}{\mathrm{d}x_i}\right) \nonumber
\eea
Therefore after $1^{st}$ iteration the new position and momentum values become,
\bea
x_i(t+dt)=\frac{1}{6}\Big[ k_1(i)+k_2(i)+k_3(i)+k_4(i) \Big] \\ 
p_i(t+dt)=\frac{1}{6}\Big[ q_1(i)+q_2(i)+q_3(i)+q_4(i)) \Big] 
\eea
We repeat this process at each time step to get the new position and momentum coordinates.

\section*{References}
\bibliographystyle{unsrt}
\bibliography{references}
\newpage
\end{document}